\documentclass[article]{biometrika}

\usepackage{amsmath}

\usepackage{times}
\usepackage{bm}
\usepackage{natbib}

\usepackage[plain,noend]{algorithm2e}

\makeatletter
\renewcommand{\algocf@captiontext}[2]{#1\algocf@typo. \AlCapFnt{}#2} 
\def\@algocf@capt@plain{top}
\renewcommand{\algocf@makecaption}[2]{%
  \addtolength{\hsize}{\algomargin}%
  \sbox\@tempboxa{\algocf@captiontext{#1}{#2}}%
  \ifdim\wd\@tempboxa >\hsize
    \hskip .5\algomargin%
    \parbox[t]{\hsize}{\algocf@captiontext{#1}{#2}}
  \else%
    \global\@minipagefalse%
    \hbox to\hsize{\box\@tempboxa}
  \fi%
  \addtolength{\hsize}{-\algomargin}%
}
\makeatother

\usepackage{amsmath}
\usepackage{amsfonts}
\usepackage{graphicx,psfrag,epsf}
\usepackage{enumerate}
\usepackage{amsgen, amstext,amsbsy,amsopn,amssymb,soul,subfigure}
\usepackage{epstopdf}

\newcommand{\cN}{{\cal N}}

\newcommand{\E}{E}
\newcommand{\I}{I}
\newcommand{\W}{W}

\newcommand{\X}{X}

\renewcommand{\P}{\mathrm{pr}}
\renewcommand{\equiv}{=}
   
\newcommand{\Var}{\mathrm{var}}

\newcommand{\Cov}{\mathrm{cov}}

\newcommand{\tm}[1]{\bar{X}_{T, #1}}
\newcommand{\cm}[1]{\bar{X}_{C, #1}}

\newcommand{\tms}[1]{\bar{X}_{T, #1}^*}
\newcommand{\cms}[1]{\bar{X}_{C, #1}^*}

\usepackage{url}
\usepackage{multirow}
\usepackage{caption} 

\begin{document}

\jname{Biometrika}

\markboth{Q. Zhou, P. A. Ernst, K. L. Morgan, D. B. Rubin \and A. Zhang }{Sequential rerandomization}

\title{Sequential rerandomization}

\author{Quan Zhou, Philip. A. Ernst}
\affil{Department of Statistics, Rice University, Houston, Texas 77005, U.S.A. \email{quan.zhou@rice.edu} \email{philip.ernst@rice.edu} }

\author{Kari Lock Morgan}
\affil{Department of Statistics, Pennsylvania State University, State College, Pennsylvania 16801, U.S.A. \email{klm47@psu.edu}}

\author{Donald B. Rubin}
\affil{Department of Statistics, Harvard University, Cambridge, Massachusetts 02138, U.S.A.
 \email{rubin@stat.harvard.edu}}

\author{\and Anru Zhang}
\affil{Department of Statistics, University of Wisconsin-Madison, Madison, Wisconsin 53706, U.S.A. 
\email{anruzhang@stat.wisc.edu}}

\maketitle

  
\begin{abstract}
The seminal work of \cite{morgan2012rerandomization}  considers rerandomization for all the units at one time.  
In practice, however, experimenters may have to rerandomize units sequentially.
For example, a clinician studying a rare disease may be unable to wait to perform an experiment until all the experimental units are recruited. 
Our work offers a mathematical framework for sequential rerandomization designs, where the experimental units are enrolled in groups. 
We formulate an adaptive rerandomization procedure for balancing treatment/control assignments over some continuous or binary covariates, using Mahalanobis distance as the imbalance measure. 
We prove in our key result, Theorem \ref{prop:compare.two.methods}, that given the same number of rerandomizations (in expected value), under certain mild assumptions, sequential rerandomization achieves better covariate balance than rerandomization at one time. 
\end{abstract}


\section{Introduction}
\label{sec.intro}
Rerandomization is a method for achieving balanced distributions of covariates across treatment groups before conducting an experiment~\citep{Hol,Urbach, Imai, morgan2012rerandomization}. 
Despite advocacy for rerandomization dating back to Sir Ronald Fisher, \citep[p.88]{Savage} a concrete mathematical foundation for rerandomization was only recently developed by the seminal work of \cite{morgan2012rerandomization}, who advise rerandomization only if ``the decision to rerandomize or not is based on a pre-specified criterion''~\citep[p.1265]{morgan2012rerandomization}. 
This work has catalyzed a surge of research in rerandomization, both theoretical and applied in nature. 
For theoretical contributions, see \cite{Ding1} and~\cite{Morgan2015}. For more applied contributions, see \cite{Imbens}, \cite{Dela} and \cite{Xu}.

The main objective of this work is to balance treatment/control assignments over some continuous (or binary) covariates by rerandomization. 
The majority of the traditional randomization procedures are developed for only discrete covariates, and continuous covariates are simply discretized by binning. 
But both the number and the boundaries of such bins are very difficult to choose, as discussed in~\cite{hu2012balancing}. 
\cite{morgan2012rerandomization}  consider rerandomization for a finite sample all recruited at one time
using Mahalanobis distance as the imbalance measure (henceforth we refer to this as  Morgan--Rubin complete rerandomization).  
The theoretical advantages of using Mahalanobis distance for continuous covariates are discussed in~\cite{rubin1979} and~\cite{greevy2004optimal}. 
When the data contains categorical covariates, as advocated by~\cite{morgan2012rerandomization},  we may combine blocking with rerandomization by applying a stratified randomization procedure to the most important categorical covaraites. 
However, in practice, a researcher may be unable to wait to perform an experiment until all experimental units can be recruited, and thus covariate-adaptive minimization methods~\citep[see][]{lin2015pursuit} might be preferred. 
To solve this problem, in this work, we consider rerandomization for sequential enrollment designs where participants arrive in groups, which we henceforth term  sequential rerandomization. 
To the best of our knowledge, a mathematical framework for sequential rerandomization has not been previously addressed.  
A unique advantage of sequential rerandomization is that while it is adaptive, it still allows for rerandomization, and thus is much less liable to selection bias than minimization procedures~\citep{berger2010minimization}.  
For more discussion on the relationship between rerandomization and other methods, e.g. finite selection model, see~\citet[Sec. 5]{morgan2012rerandomization}. 

Given the same number of rerandomizations (in expected value), a seemingly natural conjecture would be that the balance created by employing Morgan--Rubin complete rerandomization would, in expectation, be superior to that created using sequential rerandomization,  
since Morgan--Rubin complete rerandomization allows for all possible allocations of units. 
Under only mild asymptotic conditions,   
our result in Theorem \ref{prop:compare.two.methods} proves the opposite to be true (see Section~\ref{sec4}). 
The key mathematical implications for sequential rerandomization and the results needed to prove Theorem \ref{prop:compare.two.methods} are provided in Sections \ref{sec2} and \ref{sec3}. 
Section \ref{sec6} extends our results to more general settings and concludes the work with a discussion on 
optimal randomization procedures. 
All proofs and simulation studies are given in the Supplementary Material.

\section{Sequential rerandomization} \label{sec2}
Consider a sequential trial in which $2N$ units are to be divided into $K$ sequential groups, each group containing $2n_1, \dots, 2n_K$ experimental units, where $n_1+ \cdots +n_K=N$.
Let the matrix $X = ( X_1,\cdots, X_K ) \in \mathbb{R}^{p\times (2N)}$ 
be the $p$ covariates for these $2N$ units where $X_1,\dots, X_K$ are block matrices with corresponding dimensions $p \times 2n_1, \ldots, p \times 2n_K$;\, assume $X_1, \dots, X_k$ are observed sequentially.
$X$ will be treated as fixed and the sample covariance matrix of the $k$th group, denoted by
$\Cov(X_k)$ (which has dimension $p \times p$), is assumed to have rank equal to $p$.

Consider the following rerandomization procedure.
For the first group of $2n_1$ units, we randomly assign $n_1$ patients to the treatment group and the other $n_1$ to the control group. 
We denote this randomization by $W_1^{*}=(W_{1,1}^{*},...,W_{1,2n_1}^{*})^\top$, a vector of dimension $2n_1$,  
where $W^*_{1, i} = 1$ if the $i$th patient of the first group is assigned to treatment and  $W^*_{1, i} = 0$ otherwise. 
Throughout this manuscript, the superscript $*$ denotes results from a tentative allocation, subject to being accepted or rerandomized based on a specific criterion, whereas results without the superscript $*$ correspond to the actual treatment administered. 
The Mahalanobis distance between treatment and control groups corresponding to $W_1^{*}$ is
\begin{align*}
M_1^* =\;& \frac{n_1}{2}( \tms{1} - \cms{1} )^\top  \Cov(X_1)^{-1} ( \tms{1} - \cms{1} )   
\end{align*}
where $\tms{1}=  n_1^{-1}  X_1 W^*_1 $ and $\cms{1} =  n_1^{-1}  X_1 ( 1 - W^*_1 )$ are the $p$-dimensional mean vectors of the treatment (T) and control (C) groups respectively. 
This expression is based on the observation that  $\Cov (\tms{1} - \cms{1} \mid X_1  )  =  2 \Cov(X_1) / n_1 $ (see eq. (A1) in Appendix A$\cdot$2 for details).
As in \cite{morgan2012rerandomization}, we let $(\varphi_1, a_1)$ represent a pre-specified rerandomization criterion  
such that $ \varphi_1(X_1, W_1^*) = 1$ if $M_1^* <a_1$ and $0$ otherwise ($\varphi_1 = 1$ denotes an acceptable rerandomization). 
If $\varphi_1 = 0$, $W_1^{*}$ is not acceptable and the randomization is repeated; otherwise we set $W_1  \equiv W_1^*$, $M_1 \equiv M_1^*, \bar{X}_{T, 1} \equiv \tms{1}, \bar{X}_{C, 1} \equiv \cms{1} $  and proceed to consider the second group of $2n_2$ units. 
If $K=1$, we simply stop and sequential rerandomization reduces to Morgan--Rubin complete rerandomization.

The above methodology continues as follows. For the $k$th group of units, we
randomize $n_k$ units to treatment and $n_k$ units to control and denote the tentative assignment by $W_k^*$.
It should be emphasized that sequential rerandomization takes into account  all the data and fixed assignments  from the first $k-1$ groups, namely $X_{1:(k-1)}  \equiv ( X_1 , \cdots,  X_{k-1} )$ and $W_{1:(k-1)} \equiv ( W_1^\top, \dots, W_{k-1}^\top )^\top $, in addition to the data from the $k$th group.
The total number of subjects used to assess the acceptability of $W_k^{*}$ for the $k$th group is $2n_{1:k}$, where $n_{1:k} \equiv \sum_{j=1}^k n_j .$ 
The assignment of the first $k$ groups using $W_k^{*}$ is denoted by 
\begin{equation*}\label{eq:def.w1k}
W_{1:k}^* \equiv \left( W_1^\top, \dots, W_{k-1}^\top, {W_k^*}^\top \right)^\top, 
\end{equation*}
which is a vector with $2n_{1:k}$ components.   
The superscript $*$ on the right-hand side only occurs at the $k$th term because the  assignment vectors of the first $k-1$ groups are already fixed.  
The mean vectors of the first $k$ treatment and control groups are written as
\begin{align*}
   \tms{1:k}\equiv  \dfrac{1}{n_{1:k} } X_{1:k}  W_{1:k}^*  ,    & \quad
   \cms{1:k} \equiv  \dfrac{1}{n_{1:k} } X_{1:k} (1 -  W_{1:k}^*),
\end{align*}
with corresponding Mahalanobis distance for the first $k$ groups
\begin{equation}\label{eq:M_k}
M_k^* = \frac{n_{1:k}}{2}( \tms{1:k} - \cms{1:k} )^\top  \Cov(X_{1:k})^{-1} (\tms{1:k} - \cms{1:k} ), 
\end{equation}
where $\Cov(X_{1:k})$ is the sample covariance matrix of $X_{1:k}$, which is assumed to be full rank.
Given $a_k$, we decide whether $W_k^{*}$ is acceptable by evaluating the pre-specified rerandomization criterion
\begin{equation} \label{criter}
\varphi_k(X_{1:k}, W_{1:k}^* ) = \left\{\begin{array}{ll}
1 &  \quad\text{if $M_k^* <a_k$},\\
0 & \quad \text{otherwise},
\end{array}
\right.
\end{equation}
where $n_1, \ldots, n_K$ must be sufficiently large in order to ensure that an acceptable randomization can be realized. 
The threshold $a_k$ can be chosen as a function of $M_1, \dots, M_{k-1}$, but as we will see shortly in Section~\ref{sec:min.mk}, $M_{k-1}$ alone is sufficient for choosing $a_k$.
After the experimenter has concluded the sequential allocations, the Mahalanobis distance is calculated on the complete dataset $X = X_{1:K}$ using the appropriate version of~\eqref{eq:M_k}.

\section{Properties of sequential rerandomization} \label{sec3}

\subsection{Average treatment effect estimation}\label{sec:tau}
Now we present the key mathematical consequences for the sequential rerandomization framework outlined in Section \ref{sec2}.
We begin by the estimation for the true average treatment effect for the entire sample. 
Suppose the potential outcome for unit $i$ after treatment or control is $y_i(1)$ or $y_i(0)$, respectively, according to the Rubin causal model~\citep{Rubin74}. 
Let the observed response $Y_i = y_i(1)$ if $W_i = 1$ and $Y_i = y_i(0)$ otherwise. 
The average treatment effect is 
\begin{equation*}
\tau = \frac{\sum_{i=1}^{2N}y_i(1) - \sum_{i=1}^{2N} y_i(0)}{ 2N}.
\end{equation*} 
The usual estimate of $\tau$ is the difference between treatment group and control group sample means:
\begin{equation}\label{eq:def_hat_tau}
\hat{\tau} =  \bar{Y}_T - \bar{Y}_C =  \frac{1}{N}\sum_{i=1}^{2N}Y_i W_i - \frac{1}{N}\sum_{i=1}^{2N}Y_i (1-W_i)  = \dfrac{1}{N} Y^\top ( 2 W  - 1),
\end{equation}
where $Y$ is the vector of the outcomes.
As expected, $\hat{\tau}$ is an unbiased estimator for $\tau$. 
\begin{proposition}\label{th:tau_unbiased}
For our sequential rerandomization, $E \left(\hat{\tau} \mid  X, \varphi_1 = \cdots = \varphi_K = 1\right) = \tau. $
\end{proposition}
\begin{proof}
See Appendix A$\cdot$1. 
In fact, for $\hat{\tau}$ to be unbiased, we only require the rerandomization criterion satisfies $\varphi_k(X_{1:k}, W_{1:k}^*) = \varphi_k(X_{1:k}, 1 - W_{1:k}^*)$ for each $k$, and each group contains the same number of treatment and control units. 
\end{proof}

Next consider the sampling variance of $\hat{\tau}$.  
Following the argument of~\cite{morgan2012rerandomization}, when the treatment effect is an additive constant for all units, we decompose $Y_i$ as
\begin{equation}\label{eq:Y_i(W)_linear}
 Y_i = \hat{\beta}_0 + \hat{\beta}^\top X_i + \tau W_i + \hat{e}_i, \quad i=1,\ldots, 2N,
\end{equation}
where $\hat{\beta}_0 + \hat{\beta}^\top X_i$ is the projection of $y_i(0)$ onto the space spanned by $(1, {X}^\top)$, and $\hat{e}_i$ is the projection of $y_i(0)$ onto the orthogonal complement of that space. Letting $\bar{e}_T$ and $\bar{e}_C$ be the mean of $\hat{e}_i$ for the treatment and control groups respectively, by~\eqref{eq:def_hat_tau} and~\eqref{eq:Y_i(W)_linear}, we have 
\begin{equation}\label{eq:var.tau1}
\Var( \hat{\tau}  ) =  \Var \left\{ \hat{\beta}^\top (\bar{X}_T - \bar{X}_C)  + \bar{e}_T - \bar{e}_C  \right\}
=  \hat{\beta}^\top \Cov (\bar{X}_T - \bar{X}_C) \hat{\beta} +  \Var( \bar{e}_T - \bar{e}_C  ) , 
\end{equation}
where $\bar{X}_T$ and $\bar{X}_C$ are the covariate mean vectors of treatment and control groups. 
A natural line of enquiry is to find the reduction in $\Cov (\bar{X}_T - \bar{X}_C\mid X, \varphi_1 = \cdots = \varphi_k = 1)$ under sequential rerandomization relative to $\Cov (\bar{X}_T - \bar{X}_C\mid X)$ under complete randomization, which could be used to derive the reduction in the variance of the estimation for $\tau$.  
Recall that $M_K$ is the Mahalanobis distance of the entire dataset after all sequential randomized allocations have been conducted. 
\begin{theorem}\label{th:X_T-X_C_variance}  
Let $\nu = E(M_K \mid X,  \varphi_1 = \cdots = \varphi_K = 1) / p$. 
We have 
\begin{equation*}
\Cov(\bar{X}_{T} - \bar{X}_C \mid X, \varphi_1 = \cdots = \varphi_K = 1)  = \nu  \Cov(\bar{X}_{T} - \bar{X}_C \mid  X ) . 
\end{equation*}
\end{theorem}
\begin{proof}
See Appendix A$\cdot$2.
\end{proof}

\begin{theorem}\label{th:total-variance}
	Let $\tilde{\tau}$ be the estimator for $\tau$ for complete randomization and $\hat{\tau}$ be the estimator for $\tau$ for sequential rerandomization. 
	Assuming the treatment effect is additive, we have
	\begin{equation}
\dfrac{ \Var(\tilde{\tau})   - \Var(\hat{\tau})  }{ \Var(\tilde{\tau})  }  = (1-\nu) R^2,
	\end{equation}
where $R^2$ 
is the squared multiple correlation between $Y$ and $X$ in either the treatment or control group. 
\end{theorem}
\begin{proof}
See Appendix A$\cdot$3.
\end{proof}

These results can be seen as extensions of those presented in \cite{morgan2012rerandomization}. 
When $K=1$, the expression for $\nu$ reduces to eq. (9) in \cite{morgan2012rerandomization}. 
Henceforth we shall simply write  $E(M_k \mid X)$ instead of $E(M_k \mid X,  \varphi_1 = \cdots = \varphi_k = 1)$ since the notation $M_k$ clearly implies that sequential rerandomization has been conducted. 

\subsection{Asymptotic minimization of the expected Mahalanobis distance}\label{sec:min.mk}
Theorem \ref{th:X_T-X_C_variance} proves that, under the additive treatment effect model, $\Var(\hat{\tau})$ is minimized when $E(M_K \mid X)$ is minimized.
In this section we shall propose an asymptotically optimal strategy that minimizes $E(M_K \mid X)$ and thus makes the estimation of average treatment effect most precise. To this end, 
we first seek the distribution of $M_k$, whose distribution is a truncated version of the distribution of $M_k^*$. 
Recall that $X_1, \dots, X_K$ are treated as fixed and the randomness only comes from the treatment/control assignment.
We further assume the data is homogeneous so that $\Cov(X_k) \approx \Cov(X_{1:k})$. 
The heterogeneous case will be discussed in Section~\ref{sec:general}. 
By~\eqref{eq:M_k}, the distribution of $M_k^*$ depends on the $p$-dimensional random variable
\begin{equation} \label{eq5}
D_k^*   \equiv \tms{k}  - \cms{k} \equiv  \dfrac{1}{n_k} X_k W^*_k  -  \dfrac{1}{n_k} X_k (1 - W^*_k)   = 2\tms{k} - 2\bar{X}_k . 
\end{equation}
As shown below in Lemma~\ref{lm:rerandomization}, when $D^*_k$ is normally distributed,   
the distribution of $M_k^*$ is fully determined by the value of $M_{k-1}$, which is a non-central chi-squared distribution with non-centrality parameter proportional to $M_{k - 1}$. 
Consequently, when choosing the threshold $a_k$ in~\eqref{criter}, we only need to use $M_{k - 1}$, since conditional on $M_{k-1}$, $M_k^*$ is independent of $M_1, \dots, M_{k - 2}$.

\begin{lemma}
\label{lm:rerandomization} 
Assume $D_k^*   \mid  X_k  \sim \mathcal{N}\{ 0, 2 n_k^{-1} \Cov(X_k) \}$ and $\Cov(X_k) \approx \Cov(X_{1:k})$.
Let $M_{k-1}$ be the Mahalanobis distance for the first $k-1$ treatment and control groups after rerandomization with $M_0 \equiv 0$, then 
\begin{equation}\label{eq:mk.distr}
 M_k^* \mid  X_k,  M_{k-1} 
 \sim \dfrac{ n_k }{n_{1:k}  }  \chi_p^2  \left(  \dfrac{ n_{1:k} - n_k }{  n_k  }   M_{k - 1} \right), 
\end{equation}
where $\chi^2_p(\lambda)$ denotes a non-central chi-squared distribution with $p$ degrees of freedom and non-centrality parameter $\lambda$.   
\end{lemma}
\begin{proof}
See Appendix A$\cdot$4.
\end{proof}

\begin{remark}\label{re:pclt}
For sufficiently large $n_1, \dots, n_K$, the assumption that $D_k^*  \mid X_k \sim \mathcal{N}\{ 0, 2 n_k^{-1} \Cov(X_k) \}$ holds under very general settings~\citep{Ding1}.  
According to our sequential rerandomization procedure, the covariate mean of the $k$th treatment group, the term $\tms{k}$ in \eqref{eq5}, can be viewed as the mean of samples from a finite population without replacement. Under certain regularity conditions, the latter is known to follow a normal distribution asymptotically~\citep{wald1944statistical}. 
By~\cite{hoeffding1951combinatorial} and~\cite{hajek1961some}, a sufficient condition is as follows: the column vectors $X_1, \dots, X_{2N}$ are i.i.d. $p$-dimensional random vectors from a distribution with finite third absolute moments and with a positive definite covariance matrix.
Then as $n_k \rightarrow \infty$, $\sqrt{n_k} D_k^*$ converges in distribution to $\mathcal{N} \{ 0, 2 \Cov(X_k) \}$. 
This result will be used to compute $E(M_K \mid X)$ and derive the optimal strategy for sequential rerandomization. 
\end{remark}
 
Recall the sequential rerandomization criteria $\varphi_1, \dots, \varphi_K$ defined in~\eqref{criter}. 
We use the distribution given in~\eqref{eq:mk.distr} to choose $a_k$ so that $F_{M_k^*}(a_k) = \alpha_k$, where $F_{M_k^*}$ is the conditional distribution function of $M_k^*$ given $M_{k-1}$, and $\alpha_k$ is the acceptance probability of each randomization. 
The number of randomizations required for $\varphi_k$ to evaluate to 1 is distributed as a geometric random variable with expectation $s_k \equiv 1 / \alpha_k$.
Hence, if we know how to choose $s_k$,  we can choose $a_k$ accordingly by the distribution of $M_k^*$ given in Lemma~\ref{lm:rerandomization}, and we denote this by writing $a_k = a_k(M_{k-1}, s_k)$. 
Equipped with modern computational resources, it is reasonable to assume that the experimenter may perform rerandomization a very large number of times. We may therefore assume that $s_1, \dots, s_K$ are sufficiently large and that $M_1, \dots, M_K$ are correspondingly small.  
Using an asymptotic result for truncated non-central chi-squared distribution (Lemma~\ref{lm:noncentral.chi.extreme} given below), we proceed to find an asymptotic expression for the expected value of $M_k$ conditional on $M_{k-1}$ in Lemma~\ref{lm:mk.asymptotic}. 
 
\begin{lemma} \label{lm:noncentral.chi.extreme}
Let $M$ be a random variable that follows $\chi_p^2(\lambda)$ and $F_M$ be its c.d.f. As $a \downarrow 0$, 
\begin{equation*}
F_M(a)    \sim   \dfrac{  a^{p/2} e^{-\lambda/2}  } { 2^{p/2} \Gamma(p/2 + 1)  } ,   \quad \quad 
E( M \mid M < a )   \sim  \dfrac{pa}{p+2} , 
\end{equation*}
where  $\sim$ denotes asymptotic equivalence: for two positive functions $f(x)$ and $g(x)$,  we write $f \sim g$ as $x \rightarrow x_0$ if and only if $\lim_{x \rightarrow x_0} f(x) / g(x)  = 1$.
\end{lemma}
\begin{proof}
See Appendix A$\cdot$5.
\end{proof}

\begin{lemma} \label{lm:mk.asymptotic}
Suppose $M_k^* \; (k=1,\dots, K)$ follows the distribution given in Lemma~\ref{lm:rerandomization} and $\P (M^*_k < a_k \mid X_k,  M_{k-1}) = 1/s_k$.  Then as $s_k \uparrow \infty$  and $M_{k-1} \downarrow 0$, 
\begin{equation*}
E(M_k\mid  X_k,  M_{k-1}) 
 \sim  \dfrac{ n_k }{n_{1:k}  }   C_p  s_k^{-2/p} \left( 1  +   \dfrac{ n_{1:k} - n_k }{  p n_k    }   M_{k-1}  \right) ,  
 \end{equation*}
where $C_p   \equiv 2p \{ \Gamma( p/2 + 1 ) \} ^{2/p}  / (p + 2)  $. 
\end{lemma}

\begin{proof}
See Appendix A$\cdot$6.
\end{proof}

Let the expected total number of rerandomizations $S \equiv s_1 + \dots + s_K$ be sufficiently large. 
Proposition \ref{lm:best.strategy} details the asymptotically optimal strategy for choosing $s_1, \dots, s_K$, in which optimality is achieved by asymptotically minimizing $E( M_K \mid X)$ for fixed $S$.

\begin{proposition}\label{lm:best.strategy}
Suppose $M_k^* \; (k=1,\dots, K)$ follows the distribution given in Lemma~\ref{lm:rerandomization}. 
As $S \uparrow \infty$,  in order to minimize $E(M_K \mid X )$, one should choose $s_1, \dots, s_K$ so that 
\begin{equation}\label{eq:best.strategy}
s_{k-1} \approx  \left(   \dfrac{ C_p n_{k-1}}{p n_k}  s_k  \right)^{p/(p+2)},  \quad \quad 
C_p   \equiv  \dfrac{ 2p  }{ p + 2}  \Gamma( p/2 + 1 )^{2/p}. 
\end{equation}
\end{proposition} 
\begin{proof}
See Appendix A$\cdot$7.
\end{proof}

\section{Comparing sequential rerandomization with Morgan--Rubin complete rerandomization} \label{sec4}
In this section, we compare sequential rerandomization with Morgan--Rubin complete rerandomization. We begin by recalling the Morgan--Rubin complete rerandomization algorithm; $2N$ units are assumed to be enrolled when the rerandomization starts and randomizations are conducted until the Mahalanobis distance $M^*$ is smaller than some pre-specified threshold $a$, where 
\begin{equation}\label{eq:def.M}
M^* = \frac{N}{2} ( \bar{X}_{T}^* - \bar{X}_{C}^* )^\top\Cov(X)^{-1} (\bar{X}_{T}^* - \bar{X}_{C}^*).
\end{equation} 
When the rerandomization stops, let $M = M^*$. Asymptotically, the distribution of $M$ is a truncated chi-squared distribution with support $(0, a)$. 
This statistic $M$ denotes the same quantity as the statistic $M_K$ in sequential rerandomization; namely, it is the Mahalanobis distance calculated on the entire sample after all units have received treatment assignment. If (in expectation) the same number of rerandomizations are conducted in Morgan--Rubin complete rerandomization and sequential rerandomization, 
it is tempting to conjecture that $E(M \mid X)$, which we define as the expected Mahalanobis distance from Morgan--Rubin complete rerandomization, should be smaller than $E(M_K \mid X)$, 
since Morgan--Rubin complete rerandomization considers all $(2N)!/ N! N!$ 
possible allocations, whereas sequential rerandomization selects from a subset of those that are allowed by the sequential design. 
Surprisingly, as we will now prove in Theorem \ref{prop:compare.two.methods} below, under certain asymptotic conditions, the opposite holds true.

\begin{theorem}\label{prop:compare.two.methods}
Let $n_1, \dots, n_K$ be given and $S \in \mathbb{N}$ be the expected total number of rerandomizations.
For Morgan--Rubin complete rerandomization, choose the threshold $a$ such that $\P( M^* < a \mid X) = 1 / S$;
for sequential rerandomization, choose $s_1, \dots, s_K$ according to Proposition~\ref{lm:best.strategy} under the constraint $\sum_{i=1}^K s_i = S$ and then choose thresholds $a_k$ such that $\P( M_k^* < a_k \mid X_k,  M_{k-1} ) = 1 / s_k$. 
Then, assuming $M_k ^*$ given $M_{k-1}\; (k=1,\dots, K)$ follows the distribution given in Lemma~\ref{lm:rerandomization}, as $S \uparrow \infty$, 
\begin{equation}
E(M_K \mid X) \sim \dfrac{n_K}{N} E(M \mid X). 
\end{equation}
\end{theorem}

\begin{proof}
See Appendix A$\cdot$8.
\end{proof} 

\begin{corollary}\label{coro:relative.eff}
Under the assumptions of Theorem~\ref{prop:compare.two.methods} and assuming $n_1 = \dots = n_K$, as $S$ grows to infinity, $E(M_K \mid X)  \sim  E(M \mid X) / K$.
\end{corollary}
 
\begin{remark}
We pause to offer some intuition for Theorem \ref{prop:compare.two.methods}.
The rerandomization of the last group is the most important step because any imbalance between the first $K-1$ treatment and control groups may be cancelled out,  making the entire dataset balanced once again. 
Heuristically, an efficient sequential rerandomization strategy need only ensure that the imbalance accumulated in the first $K-1$ groups is sufficiently small and then perform most rerandomizations for the last group.
In fact, any strategy that satisfies the following two conditions would make Theorem~\ref{prop:compare.two.methods} hold: (i) as $S \uparrow \infty$, every $s_k$ does so too; (ii) $S \sim s_K$. 
The first condition ensures every $M_k$ will decrease to zero and thus, by Lemma~\ref{lm:rerandomization}, $N n_K^{-1}  M_K^*$ converges to a $\chi_p^2$ random variable.  
The second condition guarantees that, asymptotically, $N n_K^{-1}  M_K$ and $M$ are equivalent (in expectation) because they are truncated at the same threshold. 
\end{remark}

The consequences of the results in this section can be significant for clinical trials research.
If a large number of individuals are enrolled simultaneously, Theorem~\ref{prop:compare.two.methods} says that it is advantageous to use sequential rerandomization in lieu of Morgan--Rubin complete rerandomization. 
In Appendix B, we conduct multiple simulation studies using both simulated and real datasets to show that sequential rerandomization achieves smaller Mahalanobis distance in almost every practical setting.

\section{Discussion} \label{sec6}
\subsection{Generalizations of our results}\label{sec:general}
In practice, experimenters may prefer unequal allocation schemes where the numbers of treatment and control units are not equal~\citep{hey2014questionable}.  
Let $\omega$ be the proportion of treatment assignments. 
If $\omega$ is a constant across all the groups,  as long as we use the correct version of Mahalanobis distance (see eq. (A8) in Appendix A$\cdot$9), all our results still hold.
More generally, our results can be extended to a heterogeneous dataset where $\Cov(X_k)$ is very different across the groups. 
This happens when the clinical trial has a large time span or the groups of samples are collected at different places. 
The key is to find an appropriate form of Mahalanobis distance. 
In Appendix A$\cdot$9, we propose to standardize the data separately for each group and then compute the Mahalanobis distance using the standardized variables with proper weights (see eq. (A7)). 
It can be viewed as a generalization of the Mahalanobis distance defined in~\eqref{eq:M_k} and a corresponding generalized version of Lemma~\ref{lm:rerandomization} is also proved (see Lemma A1). 
Since Lemma~\ref{lm:rerandomization} characterizes the conditional distribution of $M^*_k$ and is the foundation of all the subsequent results, our main results, Proposition~\ref{lm:best.strategy} and Theorem~\ref{prop:compare.two.methods}, follow by the same argument.

When the data is heterogeneous, minimizing the overall imbalance may not be sufficient and thus the strategy given in Proposition~\ref{lm:best.strategy} becomes undesirable. 
For example, the effects of the covariates may change between the groups, and each group may have a unique systematic effect on the outcome. 
In such cases, one may want to achieve good balance within each group and choose a more uniform value for $(s_1, \dots, s_K)$. 
We point out that, in terms of within-group balance,  sequential rerandomization is still superior to Morgan--Rubin complete rerandomization (see Appendix A$\cdot$10 for details).  
Heuristically, this is because  even if we let the threshold $a$ of  Morgan-Rubin complete rerandomization go to zero, we are only enforcing the within-group imbalances to cancel out but their absolute values can be arbitrarily large.

\subsection{Towards an optimal procedure}\label{sec:optimal}
For classical dynamic randomization procedures, it is often assumed that the assignment of a unit must be determined as soon as he/she is enrolled. 
The seminal work of~\cite{atkinson1982optimum} proposes a type of Efron's biased coin procedure~\citep{efron1971forcing} that achieves optimum performance when the underlying model is linear~\citep[see also][]{smith1984properties}. 
A natural line of enquiry is to find the optimal procedure if the participants arrive in groups and the rerandomization technique is employed.
Such questions have to be formulated very carefully.
Even if all the participants arrive at one time, the deterministic construction that minimizes the Mahalanobis distance is usually undesirable for the following two reasons.
Firstly, for large sample sizes, the construction is not practical since finding the deterministic optimum is a non-convex optimization problem. 
Secondly, we want the procedure to possess certain degree of randomness to avoid the selection bias~\citep{antognini2017estimation}. 
\cite{qin2016optimal} introduces a procedure which can be applied when the participants arrive in pairs. 
For the $k$th pair, they consider the two 1:1 assignment schemes and choose the one that gives a smaller Mahalanobis distance of the first $2k$ units with probability $q \in (1/2, 1)$. 
In Appendix B$\cdot$3, we apply this procedure to a real dataset.
When $q = 0.75$ (the value suggested in their work), the Mahalanobis distance of the entire dataset is only sightly smaller than that of Morgan--Rubin complete rerandomization, but is greater than those of our sequential designs. 
For comparison, when $q = 1$, which makes the whole procedure deterministic, the Mahalanobis distance reduces dramatically.  
We believe that when the group size of a sequential design is small, the covariate imbalance can only be efficiently minimized at the cost of selection bias, i.e. the procedure being more deterministic.  
\cite{kapelner2014matching} offers a more complicated dynamic procedure which also uses Mahalanobis distance, but some participants may wait a long time before being assigned. 
We believe, in order to find an optimal procedure, one needs to strike a balance between the following factors: 
covariate imbalance (measured by Mahalanobis distance), group size of the sequential enrollment design, and the randomness and computational cost of the procedure.

\section*{Acknowledgement}
We thank the anonymous referees for comments which helped improve the quality of this manuscript. 
Professor Donald B. Rubin gratefully acknowledges support from the National Science Foundation, the National Institutes of Health and the Google Faculty Fellowship. 

\section*{Supplementary Material}
Appendix A includes all the proofs and some theoretical generalizations of our results. 
Appendix B provides simulation studies with both simulated and real datasets. 

\bibliographystyle{biometrika}
\bibliography{reference2}

\newpage

\begin{center}
\vspace*{0.1cm}
\large{\bf Supplementary Material}
\end{center}
 
\appendix

\appendixone
\setcounter{table}{0}
\setcounter{subsection}{0}
\setcounter{figure}{0}
\setcounter{equation}{0}
\renewcommand{\theequation}{A\arabic{equation}}
\renewcommand{\thetable}{A\arabic{table}}
\renewcommand{\thefigure}{A\arabic{figure}}

\section*{Appendix A: Proofs}
\subsection{Proof for Proposition~\ref{th:tau_unbiased}}
\begin{proof}
The proof is essentially the same as the proof for Theorem 2.1 in~\citet{morgan2012rerandomization}. For a standard completely randomized trial, $\W$ and $1-\W$ have the same probability distributions.
For sequential rerandomization, if $\varphi_k(\X_{1:k}, \W_{1:k}^*) = \varphi_k(\X_{1:k}, 1 - \W_{1:k}^*)$ for every $k$, the conditional probabilities
$$
\P (  \W_{1:k} \mid \varphi_1 = \cdots = \varphi_K= 1 ) \quad \text{and}\quad \P ( 1 - \W_{1:k} \mid \varphi_1 = \cdots = \varphi_K = 1) $$ 
are equal by symmetry. Therefore,
\begin{equation*}  
\E(W_i \mid \X, \varphi_1 = \cdots = \varphi_K = 1) = \E(1 - W_i \mid \X, \varphi_1 = \cdots = \varphi_K = 1) = \frac{1}{2}.
\end{equation*}
This gives 
\begin{equation*}
\begin{split}
& \E(\hat{\tau} \mid \X, \varphi_1  = \cdots = \varphi_k  = 1)\\
= & \dfrac{1}{N} \E \left\{  \sum_{i=1}^{2N} W_i Y_i  -  \sum_{i=1}^{2N} (1-W_i)Y_i  \mid \X, \varphi_1 =\cdots = \varphi_K = 1 \right\}\\
= & \dfrac{1}{N} \E\left\{ \sum_{i=1}^{2N}W_i y_i(1)  -  \sum_{i=1}^{2N} (1-W_i)y_i(0) \mid \X, \varphi_1 = \cdots = \varphi_K = 1\right\}\\
= & \frac{1}{2N} \sum_{i=1}^{2N} \left\{ y_i(1) - y_i(0) \right\}  = \tau.
\end{split}
\end{equation*}
The proposition is thus proved. 
\end{proof}

\subsection{Proof for Theorem~\ref{th:X_T-X_C_variance}}
\begin{proof}
The proof is similar to the proof for Theorem 3.1 in~\citet{morgan2012rerandomization}. 
Let $\Cov(\X)$ be the sample covariance matrix, which is defined as 
\begin{align*}
\left\{ \Cov(\X)\right\}_{ij} = \dfrac{1}{2N-1} \sum\limits_{k=1}^{2N}  (X_{ik} - \bar{X}_i) ( X_{jk}  - \bar{X}_j),  \quad \quad  i, j \in \{1, \dots, p\}. 
\end{align*}
Because $\bar{X}_T - \bar{X}_C = 2 (\bar{X}_T - \bar{X})$, where $\bar{X}$ is the overall mean vector of $\X$ (which is fixed), 
\begin{equation}\label{eq:cov.seq}
\Cov( \bar{X}_{T}  - \bar{X}_{C}  \mid \X ) = 4 \Cov(\bar{X}_T \mid \X) =  2 N^{-1}  \Cov(\X).
\end{equation} 
Note that $\bar{X}_T$ is the sample mean of a random subsample without replacement of size $N$ from a population of size $2N$.  
Hence, the second equality can be derived by applying eq. (2.9) of~\cite{sen1995hajek}. 
Let 
\begin{equation*}
Z = (N/2)^{1/2} \Cov(\X)^{-1/2} (\bar{X}_{T}  - \bar{X}_{C}).
\end{equation*} 
Thus $\E(Z \mid \X) = 0$ and $\Cov(Z  \mid \X ) = \I$. 
For sequential rerandomization, we may decompose $\bar{X}_T - \bar{X}_C$ into a weighted sum of $\tm{i} - \cm{i}$, where the weights are given by $n_i / N$, i.e.,
\begin{align*}
\bar{X}_T - \bar{X}_C =  \dfrac{1}{N} \sum\limits_{i=1}^K n_i \left( \tm{i} - \cm{i} \right). 
\end{align*} 
Accordingly, $Z$ can be decomposed as the sum of $K$ components, $Z^{(1)}, \dots, Z^{(K)}$, defined by 
\begin{align*}
Z^{(i)} =  \dfrac{n_i}{N} (N/2)^{1/2} \Cov(\X)^{-1/2} (\tm{i}  - \cm{i}).
\end{align*}
Given $\varphi_1 = \dots = \varphi_K = 1$, by \eqref{eq:M_k},  $Z^\top Z = M_K$.
Furthermore, if we exchange $(Z^{(1)}_i,  \dots, Z^{(K)}_i)$ with  $(Z^{(1)}_j,  \dots, Z^{(K)}_j)$, one can check that the Mahalanobis distances $M_1, \dots, M_K$ remain the same, which implies that  $Z_1, \dots, Z_p$ (the row vectors of $Z$) are exchangeable conditioning on $M_1, \dots, M_k$. Therefore,
\begin{align*}
\Var (Z_i \mid  \X,  \varphi_1 = \dots = \varphi_K = 1 )  =\;& \E(Z^\top Z 
\mid \X,  \varphi_1 = \dots = \varphi_K = 1 ) / p \\
 = \;& \E(M_K \mid \X,  \varphi_1 = \dots = \varphi_K = 1) / p. 
\end{align*}
By the property of the Mahalanobis distance, if the sign of one covariate is interchanged (e.g., $Z_i$ to $-Z_i$), $\varphi_1,\ldots, \varphi_k$ will be unchanged. 
By this symmetry, 
\begin{align*}
& \Cov (Z_i, Z_j \mid  \X,  \varphi_1 = \dots = \varphi_K = 1 ) 
=  \E (Z_i Z_j \mid  \X,  \varphi_1 = \dots = \varphi_K = 1 )\\
=\,& \Cov (-Z_i, Z_j \mid  \X,  \varphi_1 = \dots = \varphi_K = 1 ) 
=  \E (-Z_i Z_j \mid  \X,  \varphi_1 = \dots = \varphi_K = 1 ),
\end{align*}
which implies
$$\Cov (Z_i, Z_j \mid  \X,  \varphi_1 = \dots = \varphi_K = 1 ) = 0 .$$ 
Therefore, using $\bar{X}_{T}  - \bar{X}_{C} = (2/N)^{1/2} \Cov(\X)^{1/2} Z$, we obtain 
\begin{align*}
 \Cov(\bar{X}_{T} - \bar{X}_C \mid \X, \varphi_1 = \cdots = \varphi_K = 1)  
 = 2  \E(M_K \mid \X,  \varphi_1 = \dots = \varphi_K = 1 )  \Cov(\X) / Np . 
\end{align*}
Since $\Cov (\bar{X}_{T} - \bar{X}_C \mid \X) = 2 \Cov(\X)/ N$, the theorem follows. 
\end{proof}

\subsection{Proof for Theorem~\ref{th:total-variance}}
\begin{proof}
The proof is essentially the same as the proof for Theorem 3.2 in~\citet{morgan2012rerandomization}. 
By Theorem~\ref{th:X_T-X_C_variance} and~\eqref{eq:var.tau1} and assuming the normality of the mean vectors $\bar{X}_T, \bar{X}_C, \bar{e}_T, \bar{e}_C$ (so that orthogonality implies independence), 
\begin{equation*}\label{eq:var.tau}
\begin{aligned}
\Var(\tilde{\tau}) &= \hat{\beta}^\top \Cov (\bar{X}_T - \bar{X}_C \mid \X) \hat{\beta} +   \Var( \bar{e}_T - \bar{e}_C  \mid \X) , \\ 
\Var(\hat{\tau}) &= \hat{\beta}^\top \Cov (\bar{X}_T - \bar{X}_C \mid  \X, \varphi_1 = \cdots = \varphi_k = 1)\hat{\beta}
+ \Var( \bar{e}_T - \bar{e}_C  \mid  \X, \varphi_1 = \cdots = \varphi_k = 1  )    \\
&= \nu \hat{\beta}^\top \Cov (\bar{X}_T - \bar{X}_C \mid \X   ) \hat{\beta} +  \Var( \bar{e}_T - \bar{e}_C  \mid \X) . 
\end{aligned}
\end{equation*} 
Recall by~\eqref{eq:cov.seq} that $\Cov (\bar{X}_T - \bar{X}_C \mid \X) = 2 \Cov(\X) / N$.
Let $\sigma^2_e$ be the variance of $\hat{e}_1, \dots, \hat{e}_{2N}$. Using the same argument we can show that $  \Var( \bar{e}_T - \bar{e}_C  \mid \X)  = 2 \sigma_e^2 / N$. Hence,
\begin{align*}
\dfrac{ \Var(\tilde{\tau})   - \Var(\hat{\tau})  }{ \Var(\tilde{\tau})  } 
= \dfrac{ (1-\nu) \Var (\X^\top \hat{\beta} )  }{  \Var (\X^\top \hat{\beta} ) +   \sigma_e^2  } 
= (1-\nu) R^2, 
\end{align*}
which proves the theorem.
\end{proof}

\subsection{Proof for Lemma~\ref{lm:rerandomization}}
\begin{proof}
Define  $D_{1:k}^* \equiv \tms{1:k} - \cms{1:k} $. 
During the rerandomization of the $k$th group, $D_{1:k}^*$ may be decomposed as a weighted average of the constant $D_{1:(k-1)}$ (which has already been fixed in the rerandomization of the last group) and the random variable $D_k^* $ by
\begin{equation}\label{eq:decomp.Dk}
 D_{1:k}^*  = \dfrac{ n_{1:(k-1)} }{ n_{1:k} }   D_{1:(k-1)}  +  \dfrac{ n_k }{ n_{1:k} }  D_k^*.
\end{equation}
Since $D_k^* \mid \X_k \sim \mathcal{N} \{ 0, 2 n_k^{-1} \Cov(\X_k) \}$ by assumption,
\begin{align*}
  D_{1:k}^*   \mid    \X_k,  D_{1:(k-1)}  \sim \mathcal{N}  \left\{  \dfrac{ n_{1:(k-1)} }{ n_{1:k} }  D_{1: (k - 1)}  ,  \;  \dfrac{  2 n_k }{n^2_{1:k}  }  \Cov(\X_k)  \right\}. 
\end{align*} 
Hence, 
 \begin{equation*}
 \dfrac{ \sqrt{ n_{1:k} } }{ \sqrt{n_k} }  \dfrac{\sqrt{ n_{1:k} }}{\sqrt{2} } \left\{ \Cov(\X_k)\right\}^{-1/2}  D_{1:k}^*   \mid   \X_k,   D_{1:(k-1)}  
\sim \mathcal{N} \left\{ \dfrac{ n_{1:(k-1)} }{  \sqrt{2 n_k} } \left\{ \Cov(\X_k)\right\}^{-1/2} D_{1: (k - 1)} ,   \; \I  \right\}.
\end{equation*}
The squared $\ell^2$-norm of this random vector follows a non-central chi-squared distribution,  
\begin{align*}
& \dfrac{   n_{1:k} }{  n_k }   \left ( \dfrac{ n_{1:k} }{2}  D_{1:k}^{*\top} \left\{ \Cov(\X_k)\right\}^{-1}  D_{1:k}^*   \right) \mid     \X_k, D_{1:(k-1)}  \\
 \sim  \; &\chi_p^2 \left(  \dfrac{ n_{1:(k-1)} }{n_k}     \dfrac{ n_{1:(k-1)} }  {2}   D_{1:(k-1)}^\top \left\{ \Cov(\X_k)\right\}^{-1}  D_{1:(k-1)}  \right).
\end{align*}
Since we assume the samples are homogeneous so that $\Cov(\X_k) \approx \Cov(\X_{1:{k-1}}) \approx \Cov(\X_{1:k})$, by the definition of $M_k^*$ given in~\eqref{eq:M_k}, the left-hand side of the above equation is equal to $n_{1:k} M_k^* / n_k$; 
the non-centrality parameter on the right-hand side is equal to $n_{1:(k-1)} M_{k - 1} / n_k$ (note that $M_{k-1}$ has already been fixed). 
The Lemma is then verified by noting that conditioning on $D_{1:(k-1)}$ is equivalent to conditioning on $M_{k-1}$. 
\end{proof}

\subsection{Proof for Lemma~\ref{lm:noncentral.chi.extreme}}
\begin{proof}
We refer the reader to \cite{richter2000asymptotic} for more general results. 
Herein we offer a simple proof.
Let $F_{\chi_p^2}$ and $f_{\chi_p^2} $ be the c.d.f. and p.d.f. of (central) $\chi_p^2$ distribution respectively. We have 
\begin{equation}\label{eq:cdf.chi}
F_{\chi_p^2}(a) \sim       \dfrac{a^{p/2}}{  2^{p/2 }  \Gamma(p/2 + 1) },  \quad \text{ as } a \downarrow 0. 
\end{equation}
To prove this, note that by L'H\^opital's rule, 
\begin{align*}
\lim\limits_{a \downarrow 0}  \dfrac{ F_{\chi_p^2}(a) }{   a^{p/2} / 2^{p/2 }  \Gamma(p/2 + 1) }
=  \lim\limits_{a \downarrow 0} \dfrac{  f_{\chi_p^2}(a)  }{   a^{p/2 - 1} / 2^{p/2 }  \Gamma(p/2)   } = 1. 
\end{align*}
The c.d.f. of a non-central chi-squared distribution can be written as~\citep{sankaran1963approximations} 
\begin{equation}\label{eq:cdf.noncentral}
F_M(a)   = \sum\limits_{k=0}^\infty \dfrac{e^{-\lambda/2} (\lambda/2)^k }{k!}  F_{\chi^2_{p+2k}} (a) . 
\end{equation}
Using~\eqref{eq:cdf.chi}, we obtain
\begin{align*}
F_M(a) 
  \sim \dfrac{ a^{p/2} }{2^{p/2  } } \sum\limits_{k=0}^\infty \dfrac{e^{-\lambda/2} (\lambda a /4)^k }{k! \,  \Gamma(p/2 + k + 1)  } 
  \sim  \dfrac{a^{p/2} e^{-\lambda/2} }{2^{p/2  }  \Gamma(p/2 + 1)}  
\end{align*}
 as $a$ decreases to zero. 
Similarly,  using L'H\^opital's rule one can verify that 
\begin{equation*}
 \int_0^a y d F_{\chi_p^2}(y)   \sim     \dfrac{ 2 a^{p/2 + 1} }{  (p+2) 2^{p/2 }  \Gamma(p/2 )  },   \quad \text{ as } a \downarrow 0.  
\end{equation*}
Then for the non-central chi-squared distribution,  using~\eqref{eq:cdf.noncentral} we obtain
\begin{equation*}
 \int_0^a  y  d F_M (y)   \sim \dfrac{ 2 a^{p/2 + 1} e^{-\lambda/2} }{ (p+2)  2^{p/2 } \, \Gamma(p/2)   },  \quad \text{ as } a \downarrow 0, 
\end{equation*}
which leads to $\E( M \mid M < a )  =   \int_0^a  y  d F_M (y)   /F_M(a)    \sim  pa/(p + 2)  $.  
\end{proof}

\subsection{Proof for Lemma~\ref{lm:mk.asymptotic}}
\begin{proof}
Define $q_k \equiv   n_{1:k}/ n_k    $.  By Lemma~\ref{lm:rerandomization},
\begin{equation*}
q_k M^*_k \mid  \X_k,  M_{k-1}  \sim   \chi_p^2  \left\{ (q_k - 1)   M_{k - 1}  \right\}. 
\end{equation*}
By Lemma~\ref{lm:noncentral.chi.extreme}, if $\P( q_k M^*_k <  q_k a_k \mid  \X_k,  M_{k-1}  )= 1/s_k$, we have 
\begin{align*}
1/s_k  \sim  \dfrac{ (q_k a_k)^{p/2} \exp\left\{ - (q_k - 1)  M_{k-1} / 2  \right\}  }{ 2^{p/2} \Gamma(p/2 + 1)},   \quad  \text{ as } s_k \uparrow \infty
\end{align*}
which after rearrangement yields
\begin{equation*}
q_k a_k  \sim \left\{ s_k^{-1}     \exp \left( \dfrac{q_k - 1}{2}  M_{k-1}  \right)   2^{p/2} \Gamma(p/2 + 1)   \right\}^{2/p} , \quad  \text{ as } s_k \uparrow \infty. 
\end{equation*}
Employing the facts that $e^x \sim 1 + x$ and  $(1 + x)^c \sim 1 + cx$ as $x \downarrow 0$, 
we obtain 
\begin{align*}
\E( q_k M_k^* \mid  q_k M_k^* < q_k a_k ,  \X_k,  M_{k-1})  &\sim  \dfrac{2p}{p+2} \left\{ s_k^{-1}   \exp \left( \dfrac{q_k - 1}{2}  M_{k-1} \right) \Gamma(p/2 + 1)   \right\}^{2/p}   \\
 &\sim   \dfrac{2p}{p+2} \left\{ s_k^{-1}     \Gamma(p/2 + 1) \right\} ^{2/p}  \left( 1 +   \dfrac{q_k - 1}{p} M_{k-1} \right) .   
\end{align*}
By definition, $\E(  M_k^* \mid    M_k^* <  a_k ,  \X_k,  M_{k-1}) = \E(M_k \mid \X_k, M_{k-1}) $ and  the result now follows. 
 \end{proof}

\subsection{Proof for Proposition~\ref{lm:best.strategy}}
\begin{proof}
We first make two observations about the given strategy. 
Firstly, as the expected total number of rerandomizations, $S = s_1 + \cdots + s_K$, goes to infinity, for every $k$, $s_k \uparrow \infty$ (this can be easily proven by contradiction). 
Secondly, since $p/(p+2) < 1$, we have $s_{k-1} = o(s_k)$ as $s_k$ goes to infinity, i.e. $\lim_{s_k \uparrow \infty} s_{k - 1}/s_k = 0$,  for $k = 2, \dots, K$, 
which further implies that $s_1 + \cdots + s_k \sim s_k$, and in particular, $s_K \sim S$. 

By Lemma~\ref{lm:rerandomization} and by the definition of $s_1$, $a_1 = F^{-1}_{\chi_p^2} (1/s_1) \geq M_1$.  Thus we have $M_1 = o(1)$ as $s_1 \uparrow \infty$. 
(note that we do not need to write $M_1 = o_p(1)$ since $M_1 \in [0, a_1]$.)
Similarly, $a_2$ is the $s_2^{-1}$-quantile of a scaled non-central chi-squared distribution with non-centrality parameter approaching 0. Hence as $s_2 \uparrow \infty$,  $a_2 = o(1)$ and so does $M_2$.  
Using this argument iteratively, we obtain $M_k = o(1)$ for every $k$.  
Hence the condition of Lemma~\ref{lm:mk.asymptotic} is satisfied for every $k$. 
Applying Lemma~\ref{lm:mk.asymptotic} twice, we obtain
\begin{equation}\label{eq:def.g}
\begin{aligned}
\E(M_K \mid \X,  M_{K-2})  & 
=  \E \left\{  \E(M_K \mid X_k,  M_{K-1}  ) \mid X_{k - 1},  M_{K-2} \right\}  \\
& \sim \E\left\{  \dfrac{ n_K }{ N  }   C_p  s_K^{-2/p} \left( 1  +   \dfrac{ N  - n_k }{  p n_K   }   M_{K -1}  \right)  \mid X_{k - 1}, M_{K-2} \right\}  \\
& \sim \dfrac{ n_K }{ N  }   C_p  s_K^{-2/p} \left\{ 1  +   \dfrac{  n_{K-1}  }{  p n_K   }          C_p  s_{K-1}^{-2/p} \left( 1 +  \dfrac{ n_{1:(K-2)} }{p n_{K-1}}  M_{K-2} \right)  \right\} \\
& \sim \dfrac{ n_K }{ N  }   C_p  s_K^{-2/p} \left\{ 1  +   \dfrac{  n_{K-1}  }{  p n_K   }          C_p  s_{K-1}^{-2/p}      \right\}  \equiv g . 
\end{aligned}
\end{equation}
Note that in the final step we omitted the term involving $M_{K-2}$ because it is $o(1)$. 
Now consider given  $W_{1:(K-2)}$, $s_1, \ldots, s_{K-2}$, and $M_{K-2}$, how one might choose the expected number of rerandomizations of the last two groups, i.e. $s_{K-1}$ and $s_K$, to minimize $\E(M_k\mid M_{K-2})$. 
Let $s_{K}  + s_{K - 1} = \tilde{S}$. 
We can differentiate the function $g$ defined in~\eqref{eq:def.g} with respect to $s_{K-1}$ as follows
\begin{equation*}\label{eq.differentiate}
\dfrac{ \partial g  }{\partial  s_{K-1}}   \propto  ( \tilde{S} - s_{K-1} )^{ -(p+2)/p } \left\{   \dfrac{   C_p  n_{K-1}  }{  p n_K   }       s_{K-1}^{-2/p}  \left(  \dfrac{\tilde{S} }{s_{K-1} } - 2 \right) 
 -  1  \right\}.
\end{equation*}
Because $s_{K-1} = o(s_K)$ as $s_K \uparrow \infty$ for the strategy given in~\eqref{eq:best.strategy}, we have $\tilde{S}/ s_{K-1} - 2 \sim \tilde{S}/ s_{K-1}$, and 
\begin{equation*}\label{eq:deriv.g}
 \dfrac{ \partial g  }{\partial  s_{K-1}}   \sim ( \tilde{S} - s_{K-1} )^{- (p+2)/p } \left(  \dfrac{   C_p  n_{K-1}  }{  p n_K   }  \tilde{S}   / s_{K-1}^{(p+2)/p}   -  1 \right). 
\end{equation*}
The right-hand side is $0$ if 
\begin{equation*}\label{eq:best.s}
s_{K -1}  =   \left(   \dfrac{ C_p n_{K-1}}{p n_K} \tilde{S} \right)^{p/(p+2)} \sim  \left(   \dfrac{ C_p n_{K-1}}{p n_K}  s_K \right)^{p/(p+2)} , \quad \text{ as } \tilde{S} \uparrow \infty. 
\end{equation*} 
The second asymptotic equality follows from the fact that $s_{K - 1} + s_K \sim s_K$ for the given strategy. 
The same argument then can be iteratively applied to find $s_{K-2}, \dots, s_1$.  
\end{proof} 

\subsection{Proof for Theorem~\ref{prop:compare.two.methods}}
\begin{proof}
According to the strategy given in Proposition~\ref{lm:best.strategy}, 
as $S$ goes to infinity, we have, for every $k$, $s_k \uparrow \infty$ and thus $M_k \downarrow 0$. 
Hence by Lemma~\ref{lm:mk.asymptotic}, 
$$\E(M_K \mid \X)   \sim  \dfrac{ n_K }{N}  C_p  s_K^{-2/p}.$$ 
Since Morgan--Rubin complete rerandomization is equivalent to sequential rerandomization with only one group, $\E(M \mid \X) \sim C_p  S^{-2/p} $.  
But by the optimal strategy of Proposition~\ref{lm:best.strategy} we also have $s_K \sim S$ as $S \uparrow \infty$. 
The conclusion thus follows.
\end{proof}

\subsection{Extension of Lemma~\ref{lm:rerandomization} to general sequential studies}
Our results are applicable to very general sequential designs in which unequal allocation schemes are permitted and in which data is heterogeneous across the groups. 
Suppose that for the $k$th group, there are $2n_k$ units enrolled and we assign $\omega (2n_k)$ units to the treatment and $(1 - \omega) (2 n_k)$ units to the control. 
The sample covariance matrix $\Cov(\X_k)$ may be very different for different groups.
To find an appropriate expression for the Mahalanobis distance, we need to first compute the covariance matrix of $D_k^* = \tms{k} - \cms{k}$.  
Using the variance formula for the sample mean of a random subsample without replacement from a finite population (recall~\eqref{eq:cov.seq}), it is straightforward to show that
\begin{align*}
\Cov( D_k^* \mid \X_k ) = \dfrac{1}{2n_k \omega (1 - \omega)} \Cov(\X_k). 
\end{align*}
\indent Define the standardized covariate difference vector $Z^*_k$ by
\begin{equation}\label{eq:def.zk}
Z_k^* = \left\{ 2n_k \omega (1 - \omega) \right\}^{1/2} \left\{ \Cov(\X_k)  \right\}^{-1/2}  (\tms{k} - \cms{k} ). 
\end{equation}
The Mahalanobis distance between the $k$th treatment and control groups is given as $(Z_k^* )^\top Z_k^*$.
Recall that when we conduct the rerandomization of the $k$th group, the assignments of the first $k-1$ groups have already been fixed. 
Therefore, when we generate a random assignment for the $k$th group, we may compute $M_k^*$ (the Mahalanobis distance for the first $k$ groups) as
\begin{equation}\label{eq:mk.general}
M_k^* = \dfrac{1}{n_{1:k}} \left(   \sqrt{n_1}  Z_1 + \cdots  +  \sqrt{n_{k-1}}  Z_{k-1}    +  \sqrt{n_k}  Z_k^*  \right)^\top\left(   \sqrt{n_1}  Z_1 + \cdots    +  \sqrt{n_{k-1}}  Z_{k-1}    +  \sqrt{n_k}  Z_k^*  \right), 
\end{equation}
where $Z_1, \dots, Z_{k-1}$ are the corresponding realized values for the first $k - 1$ groups. 
It is straightforward to check that when $\Cov(\X_k)$ is approximately the same for every $k$, this expression for $M_k^*$ reduces to~\citep{morgan2012rerandomization, qin2016optimal}
\begin{equation}\label{eq:Mk.unequal}
M_k^* = 2n_{1:k} \omega(1 - \omega)( \tms{1:k} - \cms{1:k} )^\top  \left\{ \Cov( \X_{1:k} ) \right\}^{-1} (\tms{1:k} - \cms{1:k} ). 
\end{equation}  
If $\omega = 1/2$, (\ref{eq:Mk.unequal}) further reduces to~\eqref{eq:M_k}. Recall that by Remark~\ref{re:pclt}, as $n_k \uparrow \infty$, $Z_k^*$ asymptotically follows a standard multivariate normal distribution. 
We now prove a generalized version of Lemma~\ref{lm:rerandomization}.

\begin{lemma}[Extension of Lemma~\ref{lm:rerandomization}]
Assume that $Z_k^* \mid \X_k \sim \cN(0, \I)$ where $Z_k^*$ is defined in~\eqref{eq:def.zk}. 
Then, for the Mahalanobis distance $M_k^*$ given in~\eqref{eq:mk.general}, we have 
\begin{equation*}
 M_k^* \mid  \X_k,  M_{k-1} 
 \sim \dfrac{ n_k }{n_{1:k}  }  \chi_p^2 \left(  \dfrac{ n_{1:k} - n_k }{  n_k  }   M_{k - 1} \right), 
\end{equation*}
where $M_{k-1}$ is the realized value of the Mahalanobis distance of the first $k-1$ groups. 
\end{lemma}
\begin{proof}
By~\eqref{eq:mk.general}, 
\begin{align*}
\dfrac{ n_{1:k} }{ n_k } M_k^* = \left(   \sqrt{\dfrac{n_1}{n_k}}  Z_1 + \cdots  +  \sqrt{\dfrac{n_{k-1}}{n_k}}  Z_{k-1}    +     Z_k^*  \right)^\top  \left(   \sqrt{\dfrac{n_1}{n_k}}  Z_1 + \cdots  +  \sqrt{\dfrac{n_{k-1}}{n_k}}  Z_{k-1}    +     Z_k^*  \right). 
\end{align*}
The right-hand side is a non-central chi-squared random variable with $p$ degrees of freedom and non-centrality parameter 
\begin{align*}
 \left(   \sqrt{\dfrac{n_1}{n_k}}  Z_1 + \cdots  +  \sqrt{\dfrac{n_{k-1}}{n_k}}  Z_{k-1}     \right)^\top  \left(   \sqrt{\dfrac{n_1}{n_k}}  Z_1 + \cdots  +  \sqrt{\dfrac{n_{k-1}}{n_k}}  Z_{k-1}    \right)
  = \dfrac{n_{1:(k-1)}}{n_k} M_{k - 1}. 
\end{align*}
The lemma then follows. 
\end{proof}

\subsection{Lemma~\ref{lm:a2} and its proof} 
\begin{lemma}\label{lm:a2}
Let $Y_1, Y_2$ be two independent standard normal variables. For two positive constants $\alpha_1, \alpha_2$, we have
\begin{equation}\label{eq:lm.a2}
\E( Y_1^2  \mid  (\alpha_1 Y_1  + \alpha_2  Y_2 )^2 < c)   =  1 -  \dfrac{ 2 \beta \gamma \phi(\gamma)   }{\Phi(\gamma) - \Phi (- \gamma) }, 
\end{equation}
where 
\begin{align*}
\beta = \dfrac{\alpha_1^2}{\alpha_1^2 + \alpha_2^2},  \quad \quad \gamma = \sqrt{\dfrac{c}{\alpha_1^2 + \alpha_2^2}}. 
\end{align*}
Consequently,
\begin{align*}
\lim\limits_{c \downarrow 0}\E\left\{  Y_1^2  \mid  (\alpha_1 Y_1  + \alpha_2  Y_2 )^2 < c \right\}   = 1 - \beta. 
\end{align*}
\end{lemma}

\begin{proof}
We compute $\E( Y_1^2  I_{ \{ (\alpha_1 Y_1  + \alpha_2  Y_2 )^2 < c \} } )$, which is equal to
\begin{align*}
\int_{-\infty}^\infty  x^2 \phi(x) \left\{  \Phi\left( \dfrac{\sqrt{c} - \alpha_1 x}{\alpha_2}  \right) 
- \Phi\left( \dfrac{-\sqrt{c} - \alpha_1 x}{\alpha_2} \right) \right\} dx, 
\end{align*}
where $\phi(x), \Phi(x)$ are the p.d.f. and c.d.f. of the standard normal distribution respectively. 
Heavy calculation shows that the above display can be reexpressed as
\begin{align*}
  \Phi(\gamma) - \Phi (- \gamma) - 2 \beta \gamma \phi(\gamma) . 
\end{align*}
Since $\alpha_1 Y_1 + \alpha_2 Y _2$ is a normal variable with mean $0$ and variance $\alpha_1^2 + \alpha_2^2$, 
\begin{align*}
\P\{ ( \alpha_1  Y_1  + \alpha_2  Y_2 )^2  < c\} = \Phi(\gamma) - \Phi (- \gamma). 
\end{align*}
Eq.~\eqref{eq:lm.a2} then follows. The limit as $c$ decreases to $0$ can be computed by applying L'H\^opital's rule. 
\end{proof}

\begin{remark}
Consider the sequential rerandomization procedure in the simplest setting $p = 1$ and $K = 2$. 
As shown in~\eqref{eq:mk.general}, the Mahalanbois distances can be expressed as 
\begin{align*}
M_1 =  Z_1^2,  \quad \quad M_2 =  N^{-1} ( \sqrt{n_1} Z_1 + \sqrt{n_2} Z_2)^2, 
\end{align*}
where the corresponding proposed value $Z_k^*$ (defined in~\eqref{eq:def.zk})  asymptotically follows a standard normal distribution. When we rerandomize the first group, we are proposing $Z_1^*$ and we stop when $M_1^* < a_1$. 
Then given the value of $M_1$, we proceed to the second group and propose values for $Z_2^*$.

On the other hand, assuming the first $2n_1$ units are equally allocated to treatment and control, for Morgan-Rubin complete rerandomization, we may write
\begin{align*}
M  =  N^{-1} ( \sqrt{n_1} Z_1 + \sqrt{n_2} Z_2)^2. 
\end{align*}
Morgan-Rubin complete rerandomization proposes $Z_1^*$ and $Z_2^*$ simultaneously and stops if $N^{-1} (\sqrt{n_1}  Z_1^* + \sqrt{n_2}  Z_2^*)^2 < a$. 
By Lemma A2, when Morgan-Rubin complete rerandomization is conducted, we have $Z_1^2 \rightarrow n_2 / N$ as $a \downarrow 0$. 
That is, no matter how small the threshold $a$, we have no guarantee for the balance of the first $2n_1$ units. 
In fact, conducting a few rerandomizations using only the first $2n_1$ units would be superior.
For example, if $n_1 = n_2$, we have $Z_1^2 \rightarrow 1/2$ as $a\downarrow 0$. 
This is not surprising since as long as for both groups we perform a 1:1 assignment to treatment and control, the overall difference, $\bar{X}_T - \bar{X}_C$, will be the average of two within-group differences. As we let the threshold $a$ of  Morgan-Rubin complete rerandomization go to zero, we are only enforcing the two within-group  quantities to cancel out but their absolute values may still indeed be very large. 
However, for sequential rerandomization, since $Z_1^*$ follows $\chi_1^2$,  on average we only need less than three tries for the first group to achieve a smaller value of $M_1$. 
The same reasoning can be applied to the general case. 
\end{remark}

\newpage 

\appendixtwo
\setcounter{table}{0}
\setcounter{subsection}{0}
\setcounter{figure}{0}
\setcounter{equation}{0}
\renewcommand{\theequation}{B\arabic{equation}}
\renewcommand{\thetable}{B\arabic{table}}
\renewcommand{\thefigure}{B\arabic{figure}}

\section*{Appendix B: Simulation studies} \label{sec5}
Our simulation studies contain three parts. 
In the first,  we consider ideal datasets such that Lemma~\ref{lm:rerandomization} holds exactly; that is, $D_k^*$ (defined in~\eqref{eq5})  is exactly normally distributed.  
Then by Lemma~\ref{lm:rerandomization}, the distribution of $M_k$ is independent of $\X$ and thus we shall simply write 
$\E(M_K)$ and $\E(M)$ instead of $\E(M_K \mid \X)$ and $\E(M \mid \X)$. 
We employ Monte Carlo sampling  to compare $\E(M_K)$ with $\E(M)$ for different values of $K$ (the number of sample groups),   $p$ (the number of covariates) and $S$ (the expected total number of rerandomizations).  
From Theorem~\ref{prop:compare.two.methods}, we only know the ratio $\E(M) / \E(M_K)$ for $S \uparrow \infty$, but it is not clear whether sequential rerandomization is better for moderate $S$ as well as how quickly $\E(M) / \E(M_K)$ converges to $N/n_K$.  
These problems are investigated in Section~\ref{sec:sim1}. 

Lemma~\ref{lm:rerandomization} assumes $D_k^*$ is normally distributed for $k = 1, \dots, K$. 
When this CLT-based approximation is less accurate, either because the sample sizes are small or the distribution of $\X$ has heavy tails, there are fewer advantages of sequential rerandomization relative to Morgan--Rubin complete rerandomization (implied by Theorem \ref{prop:compare.two.methods}). 
Hence we simulate $\X$ with different sample sizes and different distributions in Section~\ref{sec:sim3} and compare the Mahalanobis distances after rerandomization from these datasets with those of the ideal datasets. 
The results confirm that the conclusions from the first simulation study are still applicable, although in some extreme cases, the advantage of sequential rerandomization becomes weaker.

Section~\ref{sec:sim2} uses a real clinical dataset with a mixture of binary and continuous covariates ($p = 12$). 
The results are consistent with those obtained for ideal datasets. 
We also use this dataset to study the effect of choosing different values of $(n_1, \dots, n_K)$. 
In all the designs considered, sequential rerandomization performs best.
Lastly, we apply the dynamic randomization method of~\cite{qin2016optimal} to this dataset. 

We now detail how $s_1, \dots, s_K$ are computed for the simulation studies.
In principle, we follow the rule specified by Proposition~\ref{lm:best.strategy}: for some given $S$, we try to find a vector $(s_1, \dots, s_K)$ such that the sum is exactly equal to $S$, and approximately, $s_{k-1} \approx  \left(    C_p n_{k-1} s_k / p n_k \right)^{p/(p+2)}$. 
However, by this rule, $s_k$ decreases exponentially as $k$ varies from $K$ to $1$. 
Thus, for the first one or for the first few groups, the value of $s_k$ that is computed according to this rule can be very small, which may greatly compromise the efficiency of sequential rerandomization. 
Thus, for a given $S$,  we set a lower bound for $s_1, \dots, s_K$. 
The value for this lower bound is somewhat arbitrary; for example, for $S \geq 2000$, we let the bound be $10$. 
For each simulation study, the value of $(s_1, \dots, s_K)$ we use will be explicitly provided.

\subsection{Study I:  ideal datasets} \label{sec:sim1}  
For the first simulation study, we assume that $n_1 = \cdots = n_K$ as well as that the dataset is ideal  so that Lemma~\ref{lm:rerandomization} holds exactly. 
We choose $p = 2, 5, 10$, $K = 3, 5, 10$, and let $S$ range from $10$ to $10,000$ (since $S$ is finite, the theoretical guarantee of optimality under Proposition~\ref{lm:best.strategy} is no longer applicable). 
For every combination of $p$ and $S$, we compute the threshold $a$ for Morgan--Rubin complete rerandomization by $a = F^{-1}_{\chi^2_p}(1/S)$ and the expected Mahalanobis distance after rerandomization by $\E(M) = p  \cdot S  \cdot  F_{\chi^2_{p+2}} (a)$. 
For sequential rerandomization, given $p, K, S$ and $n_1 = \cdots = n_K$, we compute $s_1, \dots, s_K$ by Proposition~\ref{lm:best.strategy} and we calculate the thresholds $a_1, \dots, a_K$ by Lemma~\ref{lm:rerandomization}. 
Then, using Lemma~\ref{lm:rerandomization}, we sample $(M_1, \dots, M_K)$ $100,000$ times (directly from non-central chi-squared distributions) and estimate $\E(M_K)$ by the sample average for $M_K$.

\begin{figure}[ht!]
\includegraphics[width=0.95\linewidth]{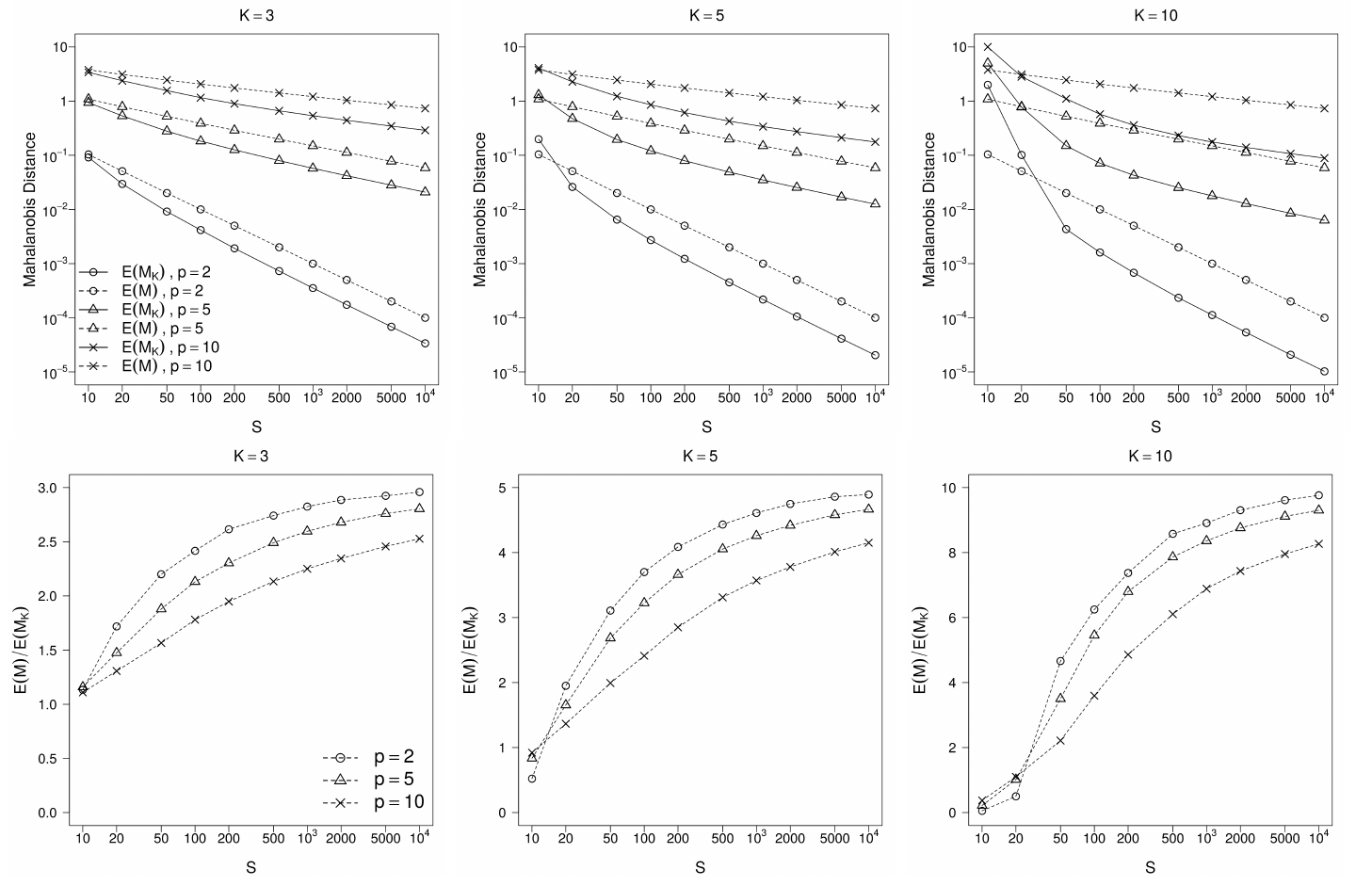}
\caption{This figure shows how the Mahalanobis distances change with increasing values of $S$, assuming that Lemma~\ref{lm:rerandomization} holds and that $n_1 = \cdots = n_K$.
 The first row of panels gives the numerical values of $\E(M)$ and $\E(M_K)$ and the second row of panels display the ratio $\E(M)/\E(M_K)$. 
$\E(M_K)$ is the average of $100,000$ Monte Carlo samples (in all cases,  SE/Mean $\leq 0.003$) and $\E(M)$ is computed exactly. 
See Table~\ref{table:s1} for the values of $s_1, \dots, s_K$.}
\label{fig:lemma1}
\end{figure}

The results are displayed in Figure~\ref{fig:lemma1}. 
Our first key observation is that $\E(M)$  is greater than $\E(M_K)$ in every case except when $S$ is extremely small; for example, $S = 10$ for $K=5$ and $S= 10, 20$ for $K = 10$.  
As $S$ increases, the ratio $\E(M) / \E(M_K)$ increases as well and eventually approaches the limit $K$. 
Next, for each $K$, the convergence of $\E(M) / \E(M_K)$ to $K$ is faster for smaller $p$. 
The main reason for this behavior is that, by Lemma~\ref{lm:mk.asymptotic}, given sufficiently small $M_{k-1}$, the expected value of $M_k$ is  $ O ( s_k^{-2/p} )$.
This behavior of $M_k$ also affects how we allocate $s_1, \dots, s_K$; recall that by Proposition~\ref{lm:best.strategy}, $s_{k-1} = O ( s_{k}^{p/(p+2)} )$. 
As a result, for larger $p$, $S$ is more evenly allocated across the groups and thus $s_K$ becomes smaller. 
Under this simulation setting, the parameter $K$ does not have a significant impact on the convergence rate of $\E(M) / \E(M_K)$, especially for $p = 2$. 
However, for a real fixed dataset,  a larger $K$ implies smaller sample size of each group, and thus $\E(M) / \E(M_K)$ cannot keep growing as $K$ grows. 

\subsection{Study II:  simulated datasets} \label{sec:sim3}
The first simulation study directly uses Lemma~\ref{lm:rerandomization}, but whether or not this lemma holds for a given dataset depends on the sample size as well as on the distribution of $\X$. 
For the second simulation study, we fix $K=5, p = 5, S = 2000$ and $n_1 = \dots = n_K$ but simulate datasets with $2n_k = 20, 50, 100$. 
For the distribution of $X$, we assume every entry $X_{ij}$ is an i.i.d. sample from some distribution $G$  and consider five choices for the distribution $G$: the standard normal distribution, the exponential distribution, the chi-squared distribution with one degree of freedom, the Weibull distribution with shape parameter $0.6$, and the log-normal distribution (exponential of a standard normal variable). 
These five distributions have an increasing excess kurtosis (the standardized fourth central moment minus three; see Table~\ref{table:distr}), which measures the tailedness; the Weibull and log-normal distributions are common examples of heavy-tailed distributions. 
Because all these distributions have finite absolute third moments,  an ideal dataset in which Lemma~\ref{lm:rerandomization} holds exactly can always be obtained by letting $N \uparrow \infty$ (see the last three rows of Table~\ref{table:distr}). 
For every combination of $2n_k$ and $G$, we perform $20,000$ Monte Carlo simulations and the data matrix $\X$ is resampled each time.
Therefore, we shall still write $\E(M_K)$ and $\E(M)$ instead of $\E(M_K \mid \X)$ and $\E(M \mid \X)$.  
It should be noted that for a simulated dataset, it is likely that $\varphi_k$ may never evaluate to 1, especially when $\X$ is heavy-tailed.
Hence, for the rerandomization of the $k$th group, we only allow at most $10 s_k$ rerandomizations and use the best assignment, i.e., the one with minimum Mahalanobis distance, if all rerandomizations fail to satisfy $\varphi_k = 1$.

\begin{table}[ht!]
\def~{\hphantom{0}}
\tbl{ $\E(M)$ and $\E(M_K)$  for different simulated datasets }{  
\begin{tabular}{llccccc}
  \multicolumn{2}{c}{Distribution of $\X$}    &   \hspace{0.3cm} $\cN(0,1)$  \hspace{0.3cm}  &  \hspace{0.3cm}  $\mathrm{Exp}$   \hspace{0.3cm}  &  \hspace{0.5cm}  $\chi_1^2$   \hspace{0.5cm} &  $\mathrm{Weibull(0.6)}$  &  $\exp( \cN(0,1) )$   \\
\multicolumn{2}{c}{Excess kurtosis}  &  0   &  6  &  12  &  37.5  &  111  \vspace{0.1cm}  \\
  & $\E(M)$  &  0.112~  & 0.112~ & 0.112~ & 0.113~ & 0.113~ \\
 $2n_k = 100$  & $\E(M_K)$  &  0.0254 & 0.0255 & 0.0255  & 0.0258 & 0.0278 \\
&  $\E(M)/\E(M_K)$ &  4.42~~ & 4.40~~ & 4.42~~ & 4.36~~   & 4.06~~   \vspace{0.1cm}   \\ 
  & $\E(M)$  &   0.112~ & 0.113~ & 0.112~ & 0.113~ & 0.113~  \\
 $2n_k = 50$ & $\E(M_K)$  &  0.0255 & 0.0256 & 0.0255  & 0.0284  & 0.0331 \\
&  $\E(M)/\E(M_K)$ &  4.41~~ & 4.40~~ & 4.41~~  & 3.97~~  & 3.42~~    \vspace{0.1cm}   \\
  & $\E(M)$  &   0.113~ & 0.113~ & 0.112~  & 0.114~ & 0.118~ \\
 $2n_k = 20$ & $\E(M_K)$  &  0.0255 & 0.0269 & 0.0324  & 0.0602 & 0.0626   \\
&  $\E(M)/\E(M_K)$ &    4.42~~    &  4.18~~ &  3.46~~   &  1.89~~   &  1.89~~  \vspace{0.1cm}  \\ 
 &  $\E(M)$  &   \multicolumn{5}{c}{ 0.112~}  \\
$2n_k = \infty$  &  $ \E(M_K) $   &   \multicolumn{5}{c}{ 0.0254} \\
 &  $ \E(M)/\E(M_K) $  &   \multicolumn{5}{c}{ 4.42~~}  
 \end{tabular}
}
\label{table:distr}
\begin{tabnote}
{\footnotesize  
For all datasets, we always use $K=5$, $p = 5$ and $S = 2000$. 
``$2n_k = \infty$"  refers to the Monte Carlo experiments using Lemma~\ref{lm:rerandomization} (see also Figure~\ref{fig:lemma1}).
As indicated in Table~\ref{table:s1}, we use $s_1 = 10, s_2 = 12, s_3 = 22, s_4 = 120, s_5 = 1836$. 
All the other estimates of $\E(M_K)$ and $\E(M)$  are obtained from $20,000$ Monte Carlo simulations. The estimates of $\E(M)$ have SE/Mean $< 0.003$ and the estimates of $\E(M_K)$  have SE/Mean $< 0.01$. 
For each distribution, the excess kurtosis is computed exactly. 
}
\end{tabnote}
\end{table}
 
The results are summarized in Table~\ref{table:distr}. 
For each choice of $2n_k$, the ratio $\E(M)/\E(M_K)$ is larger when the distribution of $\X$ has a smaller kurtosis.    
For an  ideal  dataset (or equivalently $2n_k = \infty$), we have $\E(M)/\E(M_K) = 4.42$, which is also obtained by the normal, exponential, and chi-squared distributions for $G$ when $2n_k \geq 50$. 
If $\X$ is normally distributed, $2n_k = 20$ is already sufficiently large.
For the two heavy-tailed distributions (the Weibull and log-normal distributions), the finite-sample behavior of sequential rerandomization is clearly worse than that under $2n_k = \infty$. 
Note that $2n_k = 100$ still appears to be sufficient for the Weibull$(0.6)$ distribution, but not for log-normal distribution, which has the greatest kurtosis among the five distributions.  
 
The distribution of $\X$ is often not of much practical concern.  
If $\X$ is distributed such that Lemma \ref{lm:rerandomization} does not hold, setting the thresholds $a_k$ based on Lemma \ref{lm:rerandomization} is likely to fail for large $S$.
For example, in our study for $2n_k = 20$ with $\X$ generated from a Weibull$(0.6)$ distribution, only $41$\% of the Monte Carlo simulations achieve $\varphi_k = 1$ for $k=1, \dots, K$ and the mean of $M_K$ of these samples is $0.0263$.
In practice, when $\varphi_k$ fails to evaluate to $1$, or if we can visually detect a heavy tail of the empirical distribution of some covariate, we should transform the data.

It is important to note that we did not consider discrete distributions such as the Bernoulli distribution. This is because if all covariates are binary, we would need a much larger sample size for Lemma \ref{lm:rerandomization} to hold. 
Consider the number of possible values of  $D_k^*$ given $n_k$. 
For continuous covariates, this number grows super-exponentially with $n_k$; if all covariates are binary, it only grows linearly with $n_k$ and thus converges much more slowly to a normal distribution. 
The next simulation study with real dataset shows that as long as a dataset has some continuous covariates, sequential rerandomization performs well.

\subsection{Study III:  TCGA-UCEC dataset} \label{sec:sim2}
For the last simulation study, we utilize the clinical data of TCGA-UCEC project~\citep{cancer2013integrated,ucec2016} (TCGA: The Cancer Genome Atlas; UCEC: Uterine Corpus Endometrial Carcinoma), publicly available at NCI Genomic Data Commons~\citep{grossman2016toward} and The Cancer Imaging Archive~\citep{clark2013cancer}. 
The dataset contains the clinical and demographic information of $548$ UCEC cases. 
From the original data, we choose twelve covariates that have few missing values and which are regarded as likely to be associated with the severity of tumor symptoms (more details are given in Appendix C).
For each covariate, the missing values are imputed by sampling from the observed values. 
Four covariates are continuous but only one is bell-shaped.  
Eight covariates are categorical and those with more than two levels are dichotomized, since otherwise the Mahalanobis distance 
is not practically meaningful (even though some of these covariates appear to be ordinal, it is still very difficult to attach sensible numerical values to more than two levels). 
After dichotomization,  two of the eight binary variables have the frequency of minor values less than $.1$ (we do not perform any transformation). 
Histograms for all covariates are given in Appendix C (see  Figure~\ref{fig:S1}). 
We consider five sequential enrollment designs: 
\begin{enumerate}[(i)]
\item $K=1$ and $2n_1 = 548$ (Morgan--Rubin complete rerandomization);
\item $K=3$ and $2n_1 = 184$, $2n_2 = 2n_3  = 182$;
\item $K=3$ and $2n_1 = 2n_2 = 220$, $2n_3 = 108$;
\item $K=5$ and $2n_1 = \cdots = 2n_4 = 110$, $2n_5 = 108$;
\item $K = 10$ and $2n_1 = \cdots = 2n_4 = 56 $, $2n_5 = \cdots = 2n_{10} = 54$. 
\end{enumerate}
For each design, we use $S = 2000$ and perform $20,000$ Monte Carlo simulations to obtain the sample averages for the expected  Mahalanobis distance after rerandomization.  
The entire data matrix is the same across all the repeats, but for each repeat, we resample the group labels of all the individuals (i.e. the arrival order of the individuals is resampled every time). 
For this reason, we still use the notation $\E(M_K)$ and $\E(M)$ instead of $\E(M_K \mid \X)$ and $\E(M \mid \X)$. 
As in the second simulation study, for the rerandomization of the $k$th group, we allow at most $10s_k$ rerandomizations.
 
The results are summarized in Table~\ref{table:ucec}. 
Our first observation is that the values of $\E(M_K)$ computed using the TCGA-UCEC dataset are very close to those for an ideal  dataset, which are computed using Lemma~\ref{lm:rerandomization}. 
This implies that the distribution of $D_k^*$  converges to a normal distribution quickly, despite the fact that the majority of the covariates are binary and some continuous covariates have skewed empirical distributions. 
For design (v) ($K = 10$),  the two values for $\E(M_K)$ differ most because the sample size of each group is the smallest. 
Further, although $S=2000$ is only moderate considering $p = 12$, for all designs the ratios $\E(M)/\E(M_K)$ are not far away from their limits $N/n_K$.  
Finally, design (iv) produces a larger value of $\E(M)/\E(M_K)$ than does design (iii), even though both designs have the sample size of the last group  $2n_K = 108$. 
The reason is that design (iii) has $2n_{K-1} = 220$ and design (iv) has $2n_{K-1} = 110$, which implies that design (iv) is about twice as efficient as design (iii) in minimizing the expected Mahalanobis distance of the first $440$ subjects.   
Hence, when entering the rerandomization of the last group, design (iv) tends to have a much smaller value of $M_{K-1}$ and thus a smaller $M_K$.

We conclude by applying the randomization procedure of~\cite{qin2016optimal} to this dataset. 
We first generate a random permutation of all the individuals and split them into $N = 274$ groups (each group contains $2$ individuals). 
Then for the $k$th group, we try the two possible 1:1 assignments and for each assignment we compute the Mahalanobis distance of the first $2k$ individuals; with probability $q \in (1/2, 1]$, we choose the assignment that produces a smaller Mahalanobis distance, and with probability $1 - q$, we choose the other.   
We repeat this procedure $20,000$ times. 
The sample covariance matrix of all the individuals is assumed to be known.
Using $q = 0.75$, the suggested value in \cite{qin2016optimal}, the sample mean of the Mahalanobis distance of the entire dataset after randomization is $0.926$ with standard error $4 \times 10^{-3}$, which is better than Morgan--Rubin complete rerandomization but worse than the sequential schemes we have considered. 
For comparison, using $q = 1$, the sample mean of the Mahalanobis distance of the entire dataset after randomization is $0.228$ with standard error $6 \times 10^{-4}$, which is slightly better than the sequential enrollment design (v). 
However, note that when $q = 1$, the procedure becomes deterministic once the order of all the individuals is fixed. In this case the procedure is very vulnerable to selection bias.

\begin{table}[htbp!]
\def~{\hphantom{0}}
\tbl{ $\E(M )$ and $\E(M_K )$  for TCGA-UCEC dataset}{  
\begin{tabular}{c c c c c c}
\multirow{2}{*}{ Design }  &  \multirow{2}{*}{ $N/n_K$ }  &  \multicolumn{2}{c}{TCGA-UCEC}  &   \multicolumn{2}{c}{ Ideal  dataset}  \\
  &  &   \quad  $\E(M_K)$ \quad   &  $\E(M) / \E(M_K)$    &  \quad $\E(M_K)$ \quad  &  $\E(M) / \E(M_K)$   \\
(i)  &    ~1.0 &    1.627  &     -    &    1.627  &   -  \\
(ii) &   ~3.0   &  0.723   &  2.25  &   0.723 &   2.25  \\
(iii)  &  ~5.1 &  0.539  &  3.02  &  0.536 &   3.04   \\
(iv)  &  ~5.1 &  0.455  &  3.58  &  0.453  &  3.59  \\ 
(v)  &  10.1 &  0.236  &  6.90   &  0.232 &  7.02   \\
\end{tabular}
}
\label{table:ucec}
\begin{tabnote}
{\footnotesize  
For all designs, we use $S = 2000$. 
Design (i) refers to Morgan--Rubin complete rerandomization.
Estimates of $\E(M_K)$ for TCGA-UCEC dataset are obtained from $20,000$ Monte Carlo simulations with SE/Mean $\approx 0.001$. 
``Ideal dataset"  refers to the Monte Carlo experiments using Lemma~\ref{lm:rerandomization}. 
Estimates of $\E(M_K)$ under ``Ideal dataset"  are obtained from $100,000$ Monte Carlo simulations with SE/Mean $\approx 5 \times 10^{-4}$ (exact for design (i); see Section~\ref{sec:sim1} for more details).  
See Table~\ref{table:s2} for the values of $s_1, \dots, s_K$.
}
\end{tabnote}
\end{table}

\begin{table}[htbp!]
\tbl{$(s_1, \dots, s_K)$ used in Figure~\ref{fig:lemma1}}
{\footnotesize
\begin{tabular}{  rrccc } 
 $p$ & $S$  &  \hspace{1cm} $(s_1, \dots, s_3)$  \hspace{1cm} & \hspace{1cm} $(s_1, \dots, s_5)$ \hspace{1cm}  & \hspace{1cm} $(s_1, \dots, s_{10} )$ \hspace{1cm} \vspace{0.1cm} \\
2 & 10  & (2,2,6) & (2,2,2,2,2) & (1,1,1,1,1,1,1,1,1,1) \\ 
2 & 20  & (2,4,14) & (2,2,2,3,11) & (2,2,2,2,2,2,2,2,2,2) \\ 
2 & 50  & (3,6,41) & (3,3,3,6,35) & (2,2,2,2,2,2,2,2,4,30) \\ 
2 & 100  & (5,9,86) & (4,4,4,8,80) & (3,3,3,3,3,3,3,3,7,69) \\ 
2 & 200  & (5,12,183) & (5,5,6,12,172) & (4,4,4,4,4,4,4,5,11,156) \\ 
2 & 500  & (10,22,468) & (8,8,9,20,455) & (5,5,5,5,5,5,5,6,17,442) \\ 
2 & 1000  & (10,28,962) & (10,10,11,28,941) & (8,8,8,8,8,8,8,9,26,909) \\ 
2 & 2000  & (10,37,1953) & (10,10,12,37,1931) & (10,10,10,10,10,10,10,12,36,1882) \\ 
2 & 5000  & (10,54,4936) & (10,10,13,54,4913) & (10,10,10,10,10,10,10,13,54,4863) \\ 
2 & 10000  & (10,74,9916) & (10,10,13,74,9893) & (10,10,10,10,10,10,10,13,74,9843) \\ 
5 & 10  & (2,3,5) & (2,2,2,2,2) & (1,1,1,1,1,1,1,1,1,1) \\ 
5 & 20  & (2,4,14) & (2,2,2,4,10) & (2,2,2,2,2,2,2,2,2,2) \\ 
5 & 50  & (3,8,39) & (3,3,4,8,32) & (2,2,2,2,2,2,2,3,6,27) \\ 
5 & 100  & (5,14,81) & (4,4,5,13,74) & (3,3,3,3,3,3,3,4,11,64) \\ 
5 & 200  & (5,23,172) & (5,6,8,22,159) & (4,4,4,4,4,4,4,7,19,146) \\ 
5 & 500  & (10,46,444) & (8,9,13,43,427) & (5,5,5,5,5,5,6,10,41,413) \\ 
5 & 1000  & (13,75,912) & (10,11,18,72,889) & (8,8,8,8,8,8,9,16,69,858) \\ 
5 & 2000  & (18,125,1857) & (10,12,22,120,1836) & (10,10,10,10,10,10,12,22,118,1788) \\ 
5 & 5000  & (29,244,4727) & (10,13,32,237,4708) & (10,10,10,10,10,11,13,31,235,4660) \\ 
5 & 10000  & (42,402,9556) & (10,15,43,394,9538) & (10,10,10,10,10,11,15,43,392,9489) \\ 
10 & 10  & (2,3,5) & (2,2,2,2,2) & (1,1,1,1,1,1,1,1,1,1) \\ 
10 & 20  & (2,5,13) & (2,2,2,4,10) & (2,2,2,2,2,2,2,2,2,2) \\ 
10 & 50  & (3,10,37) & (3,3,4,9,31) & (2,2,2,2,2,2,2,3,8,25) \\ 
10 & 100  & (6,19,75) & (4,4,6,16,70) & (3,3,3,3,3,3,3,6,14,59) \\ 
10 & 200  & (10,35,155) & (5,6,10,31,148) & (4,4,4,4,4,4,5,9,29,133) \\ 
10 & 500  & (19,77,404) & (8,11,19,71,391) & (5,5,5,5,5,6,7,17,68,377) \\ 
10 & 1000  & (31,139,830) & (10,14,30,129,817) & (8,8,8,8,8,9,12,28,126,785) \\ 
10 & 2000  & (50,251,1699) & (10,17,46,238,1689) & (10,10,10,10,10,11,17,45,233,1644) \\ 
10 & 5000  & (95,547,4358) & (10,24,89,525,4352) & (10,10,10,10,11,13,23,88,521,4304) \\ 
10 & 10000  & (156,984,8860) & (10,33,146,952,8859) & (10,10,10,10,11,15,32,145,949,8808) \\ 
\end{tabular}
}
\label{table:s1}
\end{table}

 \begin{table}[htbp!]
\tbl{ $(s_1, \dots, s_K)$ used in Table~\ref{table:ucec}}{  
\begin{tabular}{  c c   } 
Design  &  $(s_1, \dots, s_K)$   \\
(i)   &  (2000)  \\
(ii)  &  (62, 284, 1654) \\
(iii)  &  (94, 472, 1434) \\
(iv)  &  (10, 19, 56, 272, 1643) \\
(v)  &  (10, 10, 10, 10, 10, 12, 19, 55, 264, 1600) \\
\end{tabular}
}
\label{table:s2}
\end{table}

\clearpage
\newpage 

\appendixthree
\setcounter{table}{0}
\setcounter{subsection}{0}
\setcounter{figure}{0}
\setcounter{equation}{0}
\renewcommand{\theequation}{C\arabic{equation}}
\renewcommand{\thetable}{C\arabic{table}}
\renewcommand{\thefigure}{C\arabic{figure}}

\section*{Appendix C: Details of the TCGA-UCEC dataset}\label{sec:tcga}
\begin{figure}[htp!]
\begin{center}
\includegraphics[width=0.95\linewidth]{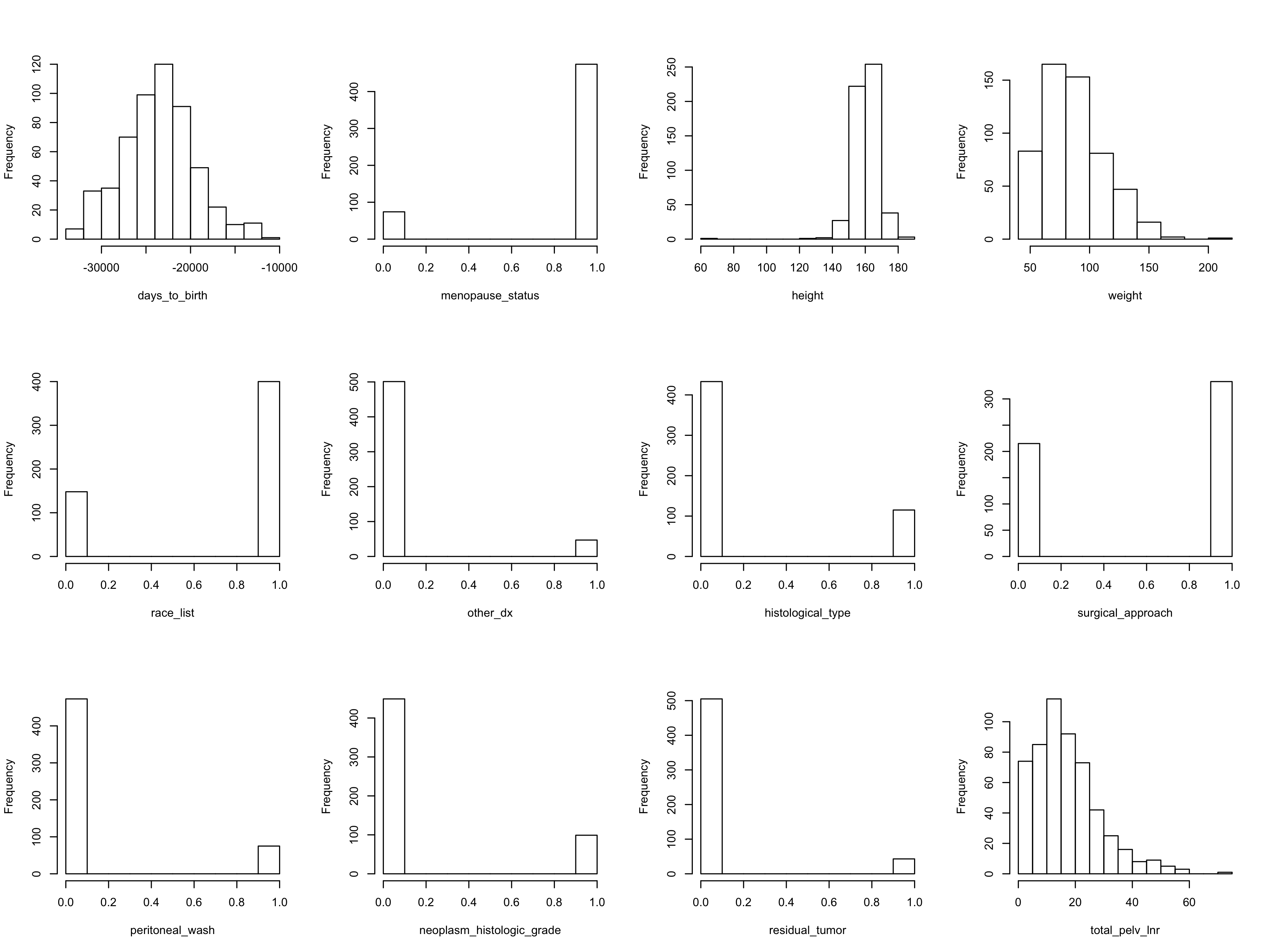}\\
\caption{Histograms for the twelve covariates of the TCGA-UCEC dataset.}\label{fig:S1}
\end{center}
\end{figure}

The TCGA-UCEC clinical dataset is available from NCI Genomic Data Commons~\footnote{\url{https://gdc.cancer.gov/}}. The dataset contains 73 covariates and 548 subjects.
The explanations of the covariates are found from the enrollment form\footnote{\url{http://www.nationwidechildrens.org/endometrial-enrollment-form}}.
From the 73 covariates, we picked the 12 covariates most likely to be associated with tumor invasion and which had few missing data: days\_to\_birth (integer), menopause\_status (categorical), height (integer), weight (integer), 
race (categorical), other\_malignancy (categorical), histological\_subtype (categorical), surgical\_approach (categorical), 
peritoneal\_wash (categorical),  tumor\_grade (categorical),  residual\_tumor (categorical), total\_pelvic\_lnr (integer, lnr: lymph nodes removed). 
Categorical covariates with more than two levels were dichotomized according to their corresponding biomedical meanings.
For each covariate, the missing values were imputed by sampling from the observed ones. 
The distributions of the twelve covariates are displayed in Figure~\ref{fig:S1}.

\end{document}